\RequirePackage{fix-cm}
\documentclass[twocolumn,epjc3]{svjour3}  
\smartqed  
\RequirePackage{graphicx}
\journalname{Eur. Phys. J. A}

\usepackage{color}
\usepackage[normalem]{ulem}

\definecolor{myGreen}{rgb}{0.2,0.72,0.2}
\renewcommand\sout{\bgroup \color[rgb]{0.55,0.00,0.99} \ULdepth=-.5ex \ULset}

\begin{document}

\title{Virtual Compton scattering at low energies with a positron beam
}


\author{Barbara Pasquini\thanksref{e1,addr1,addr2}
        \and
        Marc Vanderhaeghen\thanksref{e2,addr3} 
}

\thankstext{e1}{e-mail: barbara.pasquini@unipv.it}
\thankstext{e2}{e-mail: vandma00@uni-mainz.de}


\institute{Dipartimento di Fisica, Universit\`a degli Studi di Pavia,
               27100 Pavia, Italy
 \label{addr1}
           \and
          Istituto Nazionale di Fisica Nucleare, Sezione di Pavia,
               27100 Pavia, Italy \label{addr2}
           \and
           Institut f\"ur Kernphysik and PRISMA$^+$ Cluster of Excellence, Johannes Gutenberg Universit\"at, D-55099 Mainz, Germany \label{addr3}
}

\date{Received: date / Accepted: date}

\maketitle

\begin{abstract}
The Virtual Compton scattering (VCS) process at low energies explores the electromagnetic structure of the proton in terms of generalized polarizabilities (GPs).
 In the one-photon exchange approximation, VCS can be accessed with exclusive photon production reactions with electron or positron beams. 
The extraction of the GPs from VCS with electron beam has seen substantial progress over the past two decades. Nonetheless, a consistent picture of the GPs 
from low to higher scales demands further investigations. Complementary measurements with positron beams offer an unique  possibility, and  we present an impact study of such experimental 
program.
\end{abstract}

\section{Introduction}
\label{intro}
Virtual Compton scattering (VCS) off the proton refers to the process $\gamma^*p\rightarrow \gamma p$, where a virtual photon interacts with a proton target and a real photon is emitted in the final state. 
At low energy of the outgoing real photon, the VCS can be parametrized in terms of  generalized polarizabilities (GPs)~\cite{Guichon:1995pu,Guichon:1998xv,Drechsel:2002ar}, which are a generalization of the static polarizabilities of real Compton scattering (RCS). As the GPs depend on the photon virtuality $Q^2$, they parametrize the local polarizability response of the nucleon to a quasi-static external electromagnetic field, describing  the  spatial deformation induced in the charge, magnetization and spin densities~\cite{Gorchtein:2009qq}.
They also enter generalizations of the Gerasimov-Drell-Hearn sum rule~\cite{Gerasimov:1965et,Drell:1966jv} to finite photon virtuality, which link the GPs in model independent ways to RCS polarizabilities, moments of inclusive nucleon structure functions, and the nucleon anomalous magnetic moments and Pauli radii~\cite{Pascalutsa:2014zna,Lensky:2017dlc}. Furthermore, at low virtuality, the GPs are low-energy parameters which enter the hadronic two-photon exchange corrections, which are required in the precision extraction of the proton charge radius from muonic Hydrogen spectroscopy experiments~\cite{Carlson:2011zd,Birse:2012eb,Antognini:2013rsa,Hagelstein:2015egb}.  

The VCS process is accessed experimentally in the lepto-production of real photons, i.e., $lp\rightarrow lp\gamma$. This reaction has been investigated with an electron beam at energies below the pion production threshold and in the resonance region at MAMI~\cite{Roche:2000ng,Bensafa:2006wr,Sparveris:2008jx,Doria:2015dyx,Janssens:2008qe}, JLab~\cite{Laveissiere:2004nf,Fonvieille:2012cd} and MIT-Bates~\cite{Bourgeois:2006js,Bourgeois:2011zz}. A review of these experimental results has recently been presented  in Ref.~\cite{Fonvieille:2019eyf}. The potential of using positron beams for this reaction has not been explored so far and in this document we present a first impact study for the extraction of GPs with the proposed positron facility at JLab.

\section{Formalism and notation}
\label{sec:1}

In the $l(k)+p(p)\rightarrow l(k')+p(p')+\gamma(q')$ reaction, the final photon can be emitted either from the lepton, which corresponds to the Bethe-Heitler (BH) process, or from the proton, which is referred to as the full virtual Compton scattering (FVCS) process. The last one,  
can in turn be decomposed into a Born (B) and non-Born (NB) part.
In the Born process, the virtual photon is absorbed on the proton and the intermediate
state remains a proton, whereas the non-Born process contains all proton excitations and
meson-loop contributions. The  amplitudes of these three processes add up coherently to form the total 
$l p \rightarrow l p \gamma$ amplitude, i.e.  
\begin{eqnarray}
T^{lp\rightarrow lp\gamma}=T^{\mathrm{BH}}+T^{\mathrm{FVCS}}, \, \mathrm{with} \,\, T^{\mathrm{FVCS}}= T^{\mathrm{B}}+T^{\mathrm{NB}}.\nonumber\\
\end{eqnarray}
The BH and B terms  can be  calculated exactly
using as input only the electromagnetic form factors of the nucleon. The NB contribution contains the proton-structure  information and can be expressed with a multipole expansion in the center of mass (c.m.) frame of the $\gamma^* p$  system. In the limit of vanishing emitted photon energy, $q'_{\mathrm{cm}}\rightarrow 0$, at fixed value of the three-momentum $q_{\mathrm{cm}}$ of the virtual photon ($\gamma^*$), the multipoles which enter the low-energy expansion (LEX) of the NB part are expressed 
in terms of GPs. The leading contribution is described by six GPs~\cite{Guichon:1995pu,Drechsel:1996ag,Drechsel:1997xv}: 
two scalar GPs, i.e., the electric $\alpha_{\mathrm{E1}}(Q^2)$ and  magnetic $\beta_{\mathrm{M1}}(Q^2)$ GPs, and four spin-dependent GPs, which are functions of the photon virtuality $Q^2 \equiv -(k - k')^2$.  
The LEX formalism provides a method to analyze VCS experiments below pion production threshold in terms of structure functions which depend linearly on the GPs~\cite{Guichon:1995pu,Guichon:1998xv}. This formalism is however poorly convergent when going to higher energies. On the other hand, the dispersion relation (DR) formalism has proved 
 a very appropriate and efficient tool to analyze VCS experiments both below and above the pion production threshold~\cite{Drechsel:2002ar,Pasquini:2001yy,Pasquini:2000pk,Pasquini:2018wbl,Fonvieille:2019eyf}. The main advantage in this case is the possibility to extend the energy region of investigation, which allows for a larger sensitivity of the VCS observables to the GPs. 
For a review of the DR formalism, we refer to the original works~\cite{Pasquini:2001yy,Pasquini:2000pk,Drechsel:2002ar} and recent reviews~\cite{Pasquini:2018wbl,Fonvieille:2019eyf}, while here we only highlight its essential ingredients. In the DR approach, the VCS amplitude is decomposed in terms of 12 Lorentz invariant functions, which can be described through unsubtracted dispersion integrals relating their real and imaginary part,
at fixed values of $Q^2$ and of the four-momentum transferred $t=(p'-p)^2$, and as functions of the c.m. energy.
The high-energy behaviour of these functions guarantees the convergence of ten out of the twelve dispersion integrals.
As long as we are interested in the energy region up to the $\Delta(1232)$ resonance, the convergent integrals can be saturated by $\pi N$ intermediate  states. In practice, the calculation uses
the MAID pion photo- and electroproduction multipoles~\cite{Drechsel:1999jx,Drechsel:2007if}, which include both resonant and non-resonant production
mechanisms.
The remaining two non-converging DRs are treated as finite-energy sum rule, by cutting the dispersion integrals up to a maximum energy value and absorbing all contributions from higher energies in  asymptotic terms, which are replaced by a finite number of energy independent poles in the $t$-channel. Furthermore, in the DR model of Refs.~\cite{Pasquini:2001yy,Pasquini:2000pk},  
the dispersion integrals that show a slow convergence are supplemented by
an asymptotic contribution which parametrize the residual contribution that is not accounted for by the $\pi N$ intermediate states. 
In conclusion, besides the well known $\pi^0$ pole contribution, the asymptotic contributions can be described in terms of two phenomenological functions  that are related to the scalar GPs.
In order to provide predictions for
VCS observables, it is convenient to adopt the following parametrizations for theses phenomenological functions:
\begin{eqnarray}
\alpha_{\mathrm{E1}}(Q^2)-\alpha_{\mathrm{E1}}^{\pi N}(Q^2)&=&(\alpha_{\mathrm{E1}}^{\mathrm{exp}}-\alpha_{\mathrm{E1}}^{\pi N})f_\alpha(Q^2),\label{alpha-fit}\\
\beta_{\mathrm{M1}}(Q^2)-\beta_{\mathrm{M1}}^{\pi N}(Q^2)&=&(\beta_{\mathrm{M1}}^{\mathrm{exp}}-\beta_{\mathrm{M1}}^{\pi N})f_\beta(Q^2),\label{beta-fit}
\end{eqnarray}
where $\alpha_{\mathrm{E1}}$ and $\beta_{\mathrm{M1}}$ are the RCS polarizabilities, with  superscripts exp and $\pi N$ indicating,
respectively, the experimental value~\cite{Tanabashi:2018oca} and the $\pi N$ contribution evaluated from unsubtracted DRs.
In Eqs.~(\ref{alpha-fit}) and (\ref{beta-fit}), $f_\alpha(Q^2)$ and $f_\beta(Q^2)$ are fit functions, with
the constraints $f_\alpha(0)=f_\beta(0)=1$. Their functional form is unknown and should be adjusted by a fit to the experimental
cross sections. However, in order to provide predictions for VCS observables, it is convenient to adopt the following dipole parametrizations
\begin{eqnarray}
f_\alpha(Q^2)=\frac{1}{(1+Q^2/\Lambda_\alpha^2)^2},\; f_\beta(Q^2)=\frac{1}{(1+Q^2/\Lambda_\beta^2)^2},
\label{eq:asymp}
\end{eqnarray}
where the $\Lambda_\alpha$ and $\Lambda_\beta$ are free parameters, not necessarily constant
with $Q^2$, which can be adjusted by a fit to the experimental cross sections.
\begin{figure*}[h]
\begin{center}
  \includegraphics[width=0.65\textwidth]{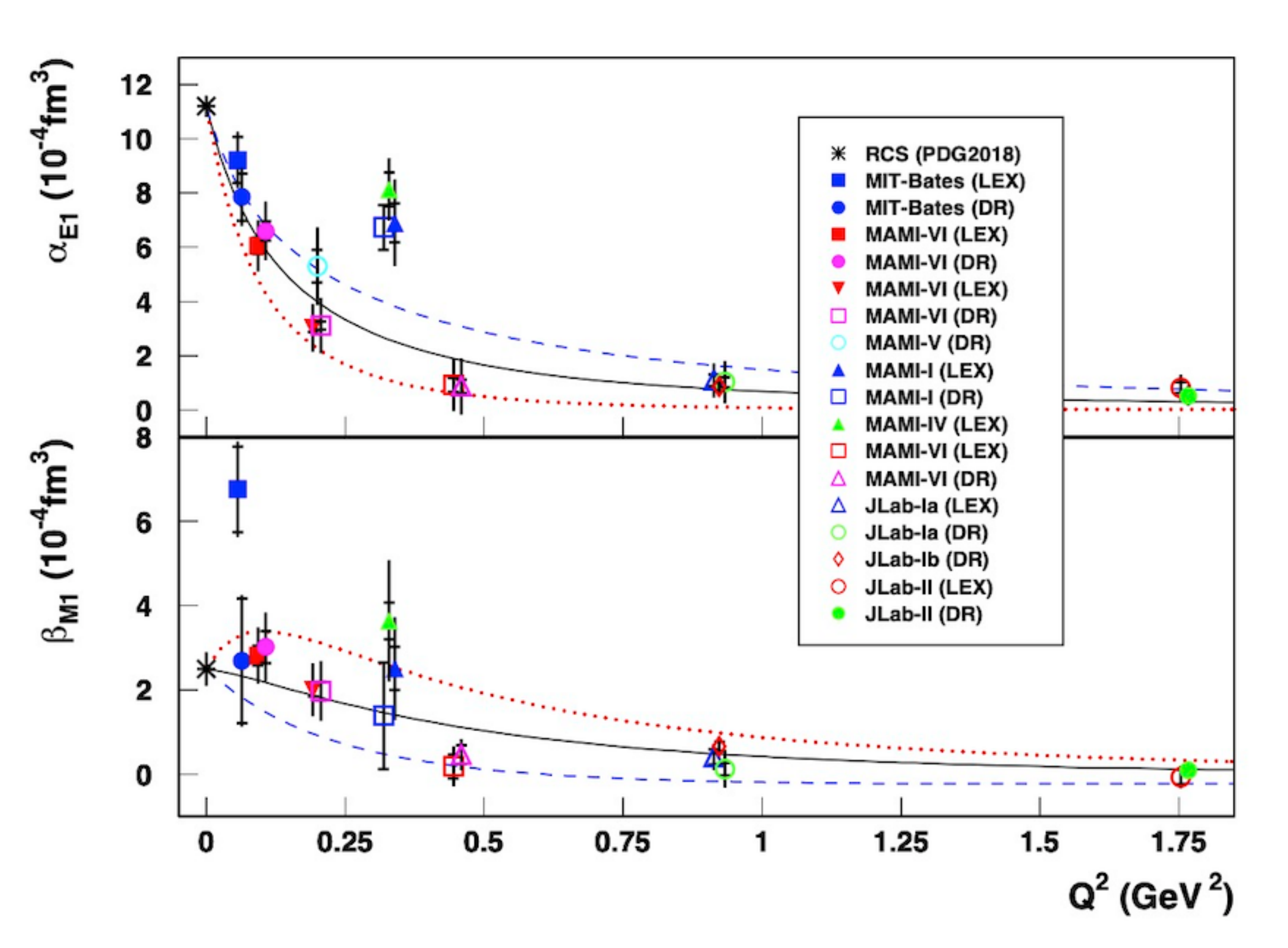}
  \caption{World data for the electric and magnetic GPs of the proton as a function of $Q^2$, including
the RCS point~\cite{Tanabashi:2018oca}. Inner (outer) error bars are statistical (total). Some points are slightly shifted for visibility. The results are from the DR calculation for three
different values  of the dipole mass parameters:  $\Lambda_\alpha= \Lambda_\beta$ 0.5 GeV (red dotted curves); $\Lambda_\alpha= \Lambda_\beta$ 0.7 GeV (black solid curves) and $\Lambda_\alpha= \Lambda_\beta$ 0.9 GeV (blue dashed curves).}
\label{fig:1}       
\end{center}
\end{figure*}

In Fig.~\ref{fig:1}, we show the world data for the GP extractions in comparison with the DR calculation for different values of the dipole mass parameters in Eq.~(\ref{eq:asymp}).
The experimental extractions from VCS measurements below pion production threshold have been obtained by applying both the LEX formalism and the DR approach. In all these analyses, the two methods give consistent results, except for the MIT-Bates measurement  in the in-plane kinematics where the LEX at leading order was shown not to be applicable.
The  experimental extractions are performed separately at each $Q^2$ value,
without making any assumption on the global $Q^2$ dependence. Yet, 
the dipole Ansatz in Eq.~(\ref{eq:asymp}) with $\Lambda_\alpha=\Lambda_\beta=0.7$ GeV   reproduces well the smooth fall-off in $Q^2$ of the data. The only exception are the MAMI-I and MAMI-IV data points at $Q^2=0.33$ GeV$^2$, mainly for the electric GP.  New data from the JLab VCS experiment E12-15-001 are currently exploring the intermediate $Q^2$ range of $0.3-0.77$ GeV$^2$ and are expected to help in the understanding of this puzzling behavior~\cite{Sparveris:2016}. 

Further insights could be gained by exploiting  VCS measurements with positron beams in the same kinematics.  As a matter of fact,  the combination of data with positron and electron  beams gives  access to various asymmetries that would allow us to better constrain the scalar GPs, as discussed in detail  in Sect.~\ref{sec:asymm}.

\section{VCS asymmetries using electron and positron beams}
\label{sec:asymm}

The kinematics of the $lp\rightarrow lp\gamma$ reaction are defined by five independent variables. We will adopt the set of variables $(E_{\mathrm{lab}},Q^2, W,\theta,\phi)$, with $E_{\mathrm{lab}}$ the incoming electron energy in the lab frame, $W$ the c.m. energy of the $\gamma^*p\rightarrow \gamma p$  subprocess, $\theta$ the polar angle (in the $\gamma^*p$ c.m. frame) between the initial and final photon, and $\phi$ the azimuthal angle between the leptonic and hadronic planes. The reference plane ($\phi = 0^o$) corresponds to the plane spanned by both lepton directions.

The lepton beam charge ($e$) and polarization ($\lambda$) dependence of the $l p \rightarrow lp \gamma$ differential cross section is given by
\begin{eqnarray}
d\sigma^e_\lambda=d\sigma_{\mathrm{BH}}+d\sigma_{\mathrm{VCS}}+\lambda \,d\tilde{\sigma}_{\mathrm{VCS}}+e\,(d\sigma_{\mathrm{INT}}+\lambda \,d\tilde{\sigma}_{\mathrm{INT}}),\nonumber\\
\end{eqnarray}
where $d\sigma$ ($d\tilde{\sigma}$) are the polarization independent (dependent) contributions which are even (odd) functions 
 of the azimuthal angle $\phi$.
$d\sigma_{\mathrm{INT}}$ involves the real part of the $\mathrm{VCS}$ amplitude that contains the GP effects, while $d\tilde{\sigma}_{\mathrm{INT}}$ is proportional to the imaginary part of the $\mathrm{VCS}$ amplitude which does not depend on the GPs. 
Combining lepton beams of opposite charge
and different polarization provides a complete separation of
the four unknown $\mathrm{INT}$ and $\mathrm{VCS}$ contributions. In the following, we will discuss the kinematical conditions that have been explored in the recent VCS measurements at JLab using an electron beam, to assess the impact of using complementary measurements with a positron beam.

Using unpolarized electron and positron beams, we can construct the unpolarized beam-charge asymmetry (BCA) $A^C_{UU}$ as
\begin{eqnarray}
A^C_{UU}&=&\frac{(d\sigma ^+_++d\sigma^+_-)-(d\sigma^-_++d\sigma^-_-)}{d\sigma^+_++d\sigma^+_-+d\sigma^-_++d\sigma^-_-}\nonumber\\
&=&\frac{d\sigma_{\mathrm{INT}}}{d\sigma_{\mathrm{BH}}+d\sigma_{\mathrm{VCS}}}.
\end{eqnarray}

With polarized lepton beams, we can construct the lepton beam-spin asymmetry (BSA)
\begin{eqnarray}
A^e_{LU}&=&\frac{d\sigma^e_+-d\sigma^e_-}{d\sigma^e_++d\sigma^e_-}\nonumber\\
&=&\frac{ d\tilde{\sigma}_{\mathrm{VCS}} +e d\tilde{\sigma}_{\mathrm{INT}}}{d\sigma_{\mathrm{BH}}+d\sigma_{\mathrm{VCS}} +e\,d\sigma_{\mathrm{INT}}  }.
\end{eqnarray}

One can combine the observable asymmetries $A^C_{UU}$, $A^+_{LU}$, and $A^-_{LU}$ to separate the contribution from the $d\tilde{\sigma}_{\mathrm{VCS}}$ and $d\tilde{\sigma}_{\mathrm{INT}}$ terms as:
\begin{eqnarray}
\tilde A_{\mathrm{VCS}} &\equiv& A^+_{LU} \left( 1 + A^C_{UU} \right) 
+ A^-_{LU} \left( 1 - A^C_{UU} \right) \nonumber \\
&=&\frac{2 d\tilde{\sigma}_{\mathrm{VCS}}}{d\sigma_{\mathrm{BH}}+d\sigma_{\mathrm{VCS}}  },
\label{eq:combplus}
\end{eqnarray}
and
\begin{eqnarray}
\tilde A_{\mathrm{INT}} &\equiv& A^+_{LU} \left( 1 + A^C_{UU} \right) 
- A^-_{LU} \left( 1 - A^C_{UU} \right) \nonumber \\
&=&\frac{2 d\tilde{\sigma}_{\mathrm{INT}}}{d\sigma_{\mathrm{BH}}+d\sigma_{\mathrm{VCS}}  }.
\label{eq:combminus}
\end{eqnarray}

The asymmetry combination $\tilde A_{\mathrm{VCS}}$ of Eq.~(\ref{eq:combplus}) is proportional to the imaginary part of the interference between the VCS amplitudes with either transverse or longitudinal virtual photon polarization. In the $\Delta(1232)$ resonance region, where the imaginary part of the transverse amplitude dominates by far over the imaginary part of the longitudinal amplitude, this combination then basically originates from the interference product of the imaginary part of the transverse VCS amplitude due to $\Delta(1232)$ excitation, and the real part of the longitudinal VCS amplitude. The latter amplitude has the sensitivity to the GP $\alpha_{\mathrm{E1}}(Q^2)$.  
On the other hand, the asymmetry combination 
$\tilde A_{\mathrm{INT}}$ of Eq.~(\ref{eq:combminus}) is proportional to the interference between the real BH amplitude and the imaginary VCS amplitude. As the latter does not depend on the GPs, it provides a cross-check of the unitarity input in the dispersive formalism. 

\section{Results and discussion}
\label{sec:results}

\begin{figure*}[t]
\begin{center}
 \includegraphics[width=0.42\textwidth]{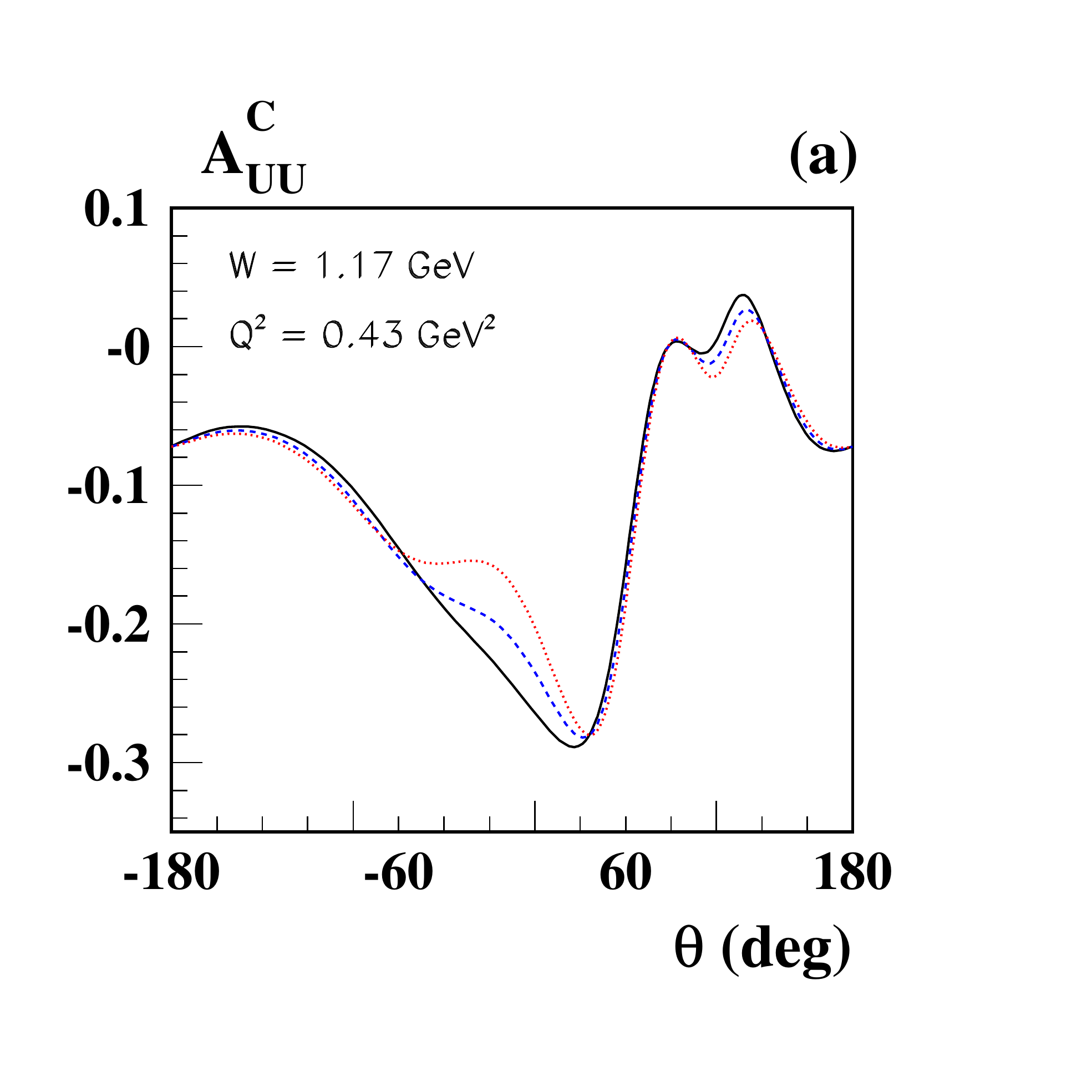}
 \includegraphics[width=0.42\textwidth]{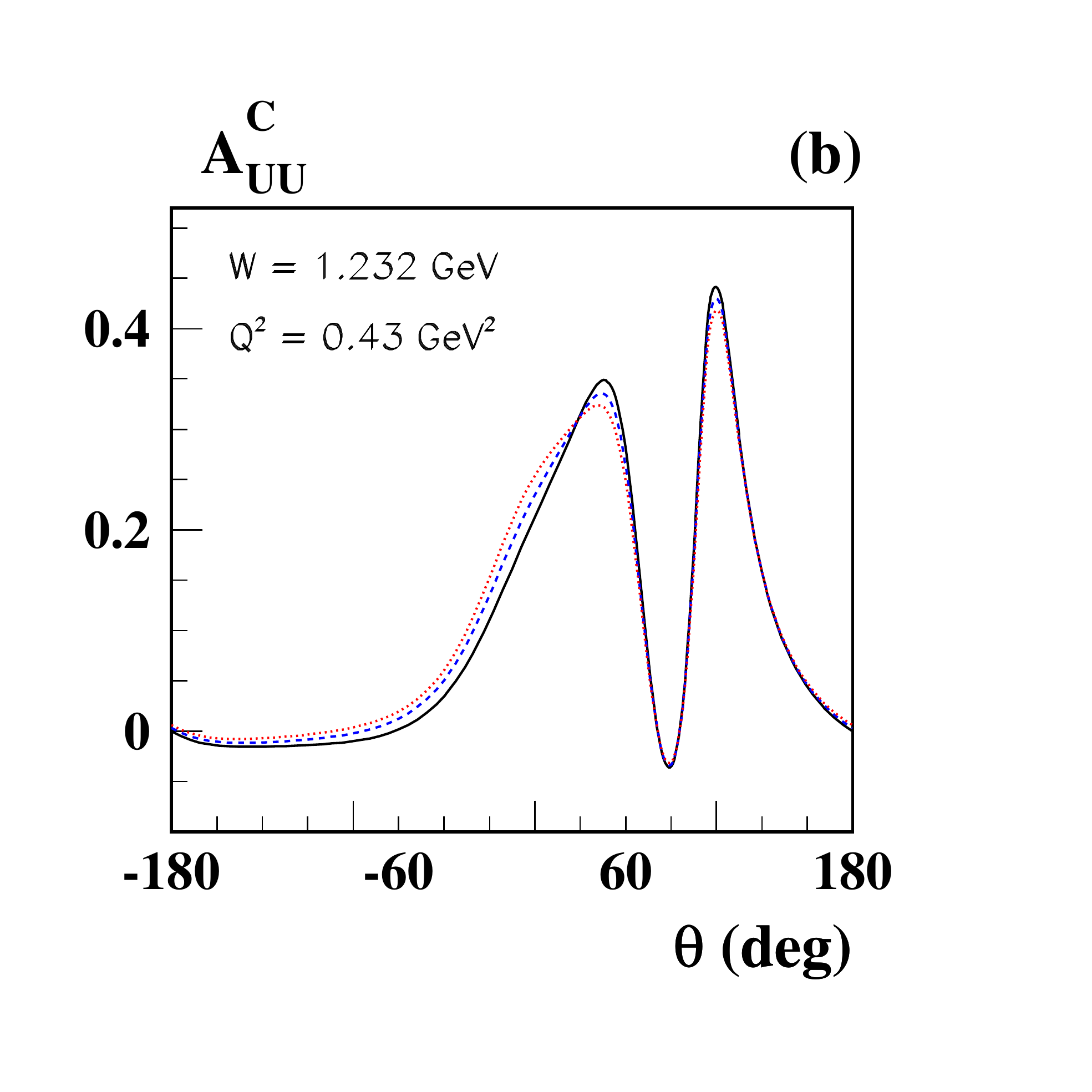}
\\
\vspace{-10 mm}
 \includegraphics[width=0.42\textwidth]{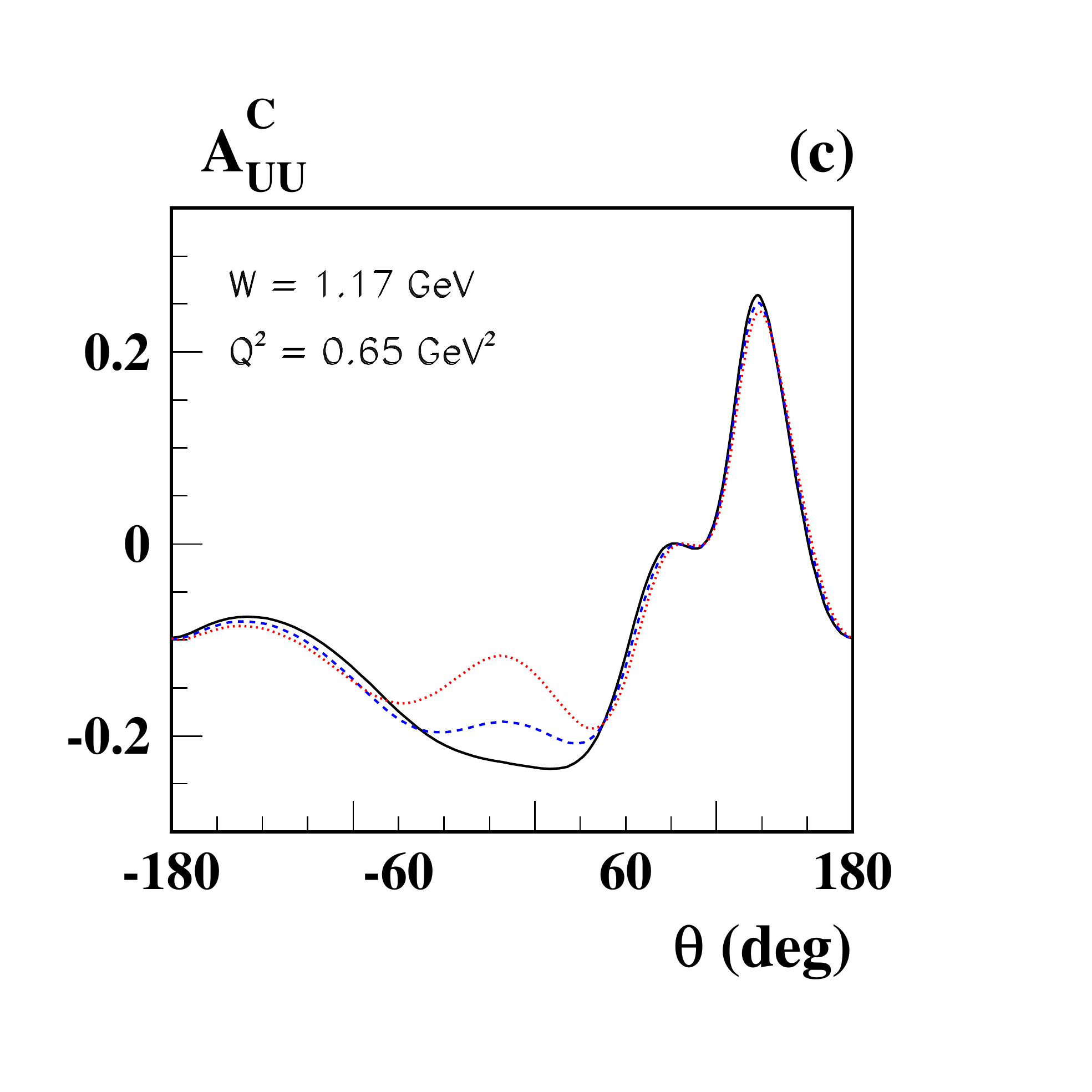}
 \includegraphics[width=0.42\textwidth]{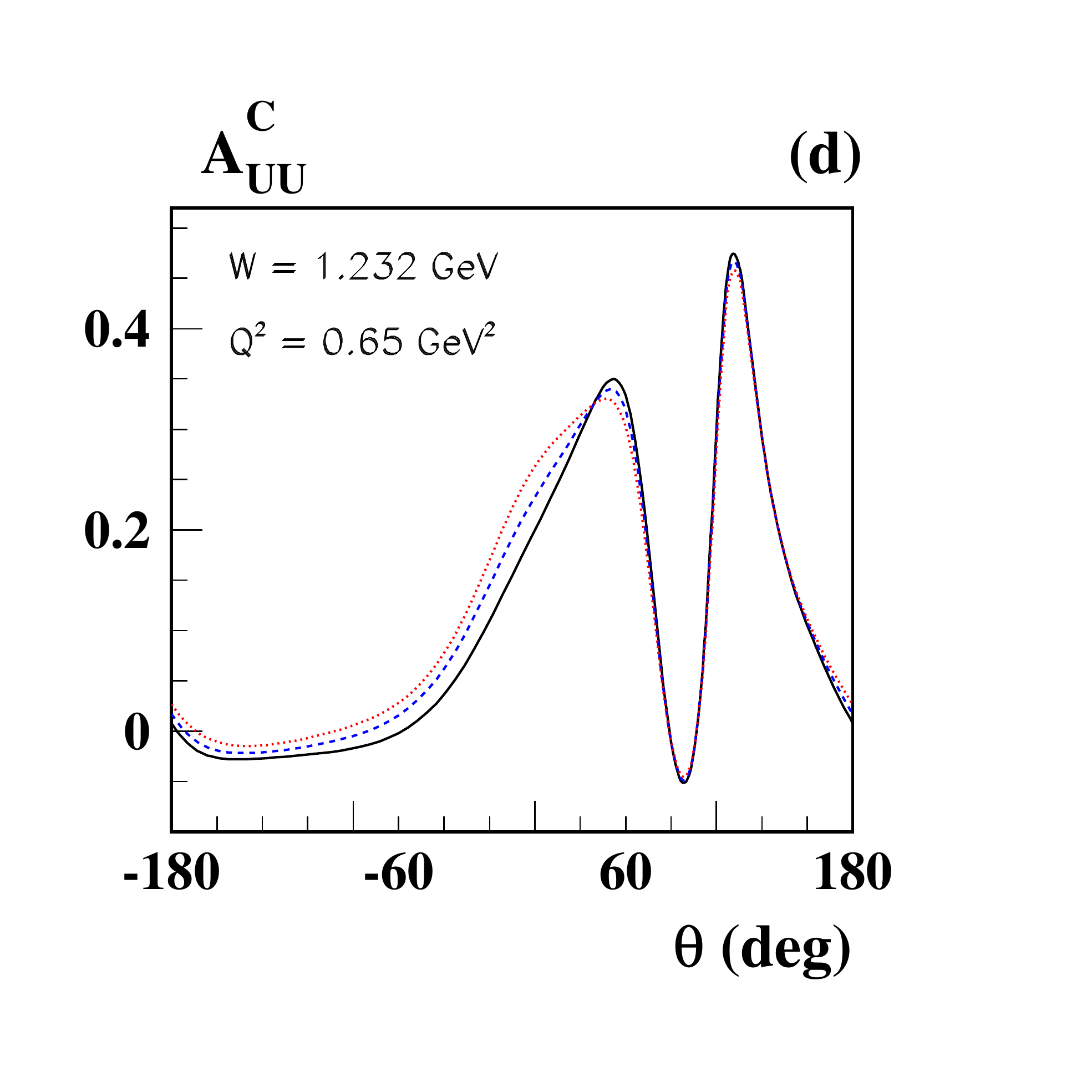}
 \end{center}
    \caption{The beam-charge asymmetry $A_{UU}^C$ as function of the photon scattering angle for in-plane kinematics, for  c.m. energy $W=1.17$ GeV (panels (a) and (c)) and $W=1.232$ GeV (panels (b) and (d)) at different values of the four-momentum transfer: $Q^2=0.43$ GeV$^2$ (panels (a) and (b)) and $Q^2=0.65$ GeV$^2$ (panels (c) and (d)). The results are for $\Lambda_\alpha=\Lambda_\beta=0.5$ GeV (red dotted curves), $\Lambda_\alpha=\Lambda_\beta=0.7$ (black solid curves) and $\Lambda_\alpha=\Lambda_\beta=0.9$ GeV (blue dashed curves). }
\label{fig:2}       
\end{figure*}

In this section, we explore for a positron beam the same kinematical range which has been measured by the E12-15-001 experiment at JLab~\cite{Sparveris:2016} using an electron beam. The experiment utilized the SHMS and HMS spectrometers \cite{SHMS,HMS} in Hall C to detect, respectively, electrons and protons in coincidence, while the reconstructed missing mass has been used for the identification of the photon. An electron beam of energy $E_{\rm lab}$ = 4.55 GeV and a 10 cm liquid hydrogen target were employed for the measurements. 

\subsection{Beam-charge asymmetry}

In Fig.~\ref{fig:2}, we show the results for the charge asymmetry for in-plane kinematics as function of the photon scattering angle (the negative and positive values correspond to $\phi=180^o$ and $\phi=0^o$, respectively) at two different values of the four-momentum transfer $Q^2$ for the center of mass energy of $W=1.17$ GeV and $W=1.232$ GeV. 
The calculations correspond to a variation of the mass-scale parameters in the range $\Lambda_\alpha=\Lambda_\beta=0.7\pm 0.2 $ GeV, which is the same range as shown along with the GP world data on Fig.~\ref{fig:1}. One notices from Fig.~\ref{fig:2} that the sensitivity to the GPs is mainly at lower energies in the forward region $-60^o\le \theta\le 40^o$  and it slightly increases at higher values of $Q^2$. 
In the region where the BH process is very large, which corresponds roughly with the angular range $70^o\le \theta\le 100^o$ in Fig.~\ref{fig:2}, one notices that the beam-charge asymmetry becomes very small as the cross section for the BH process by itself is identical between electron and positron beams. In the direction of the BH peaks the asymmetry has zeros. 

\begin{figure*}[h!]
\begin{center}
 \includegraphics[width=0.40\textwidth]{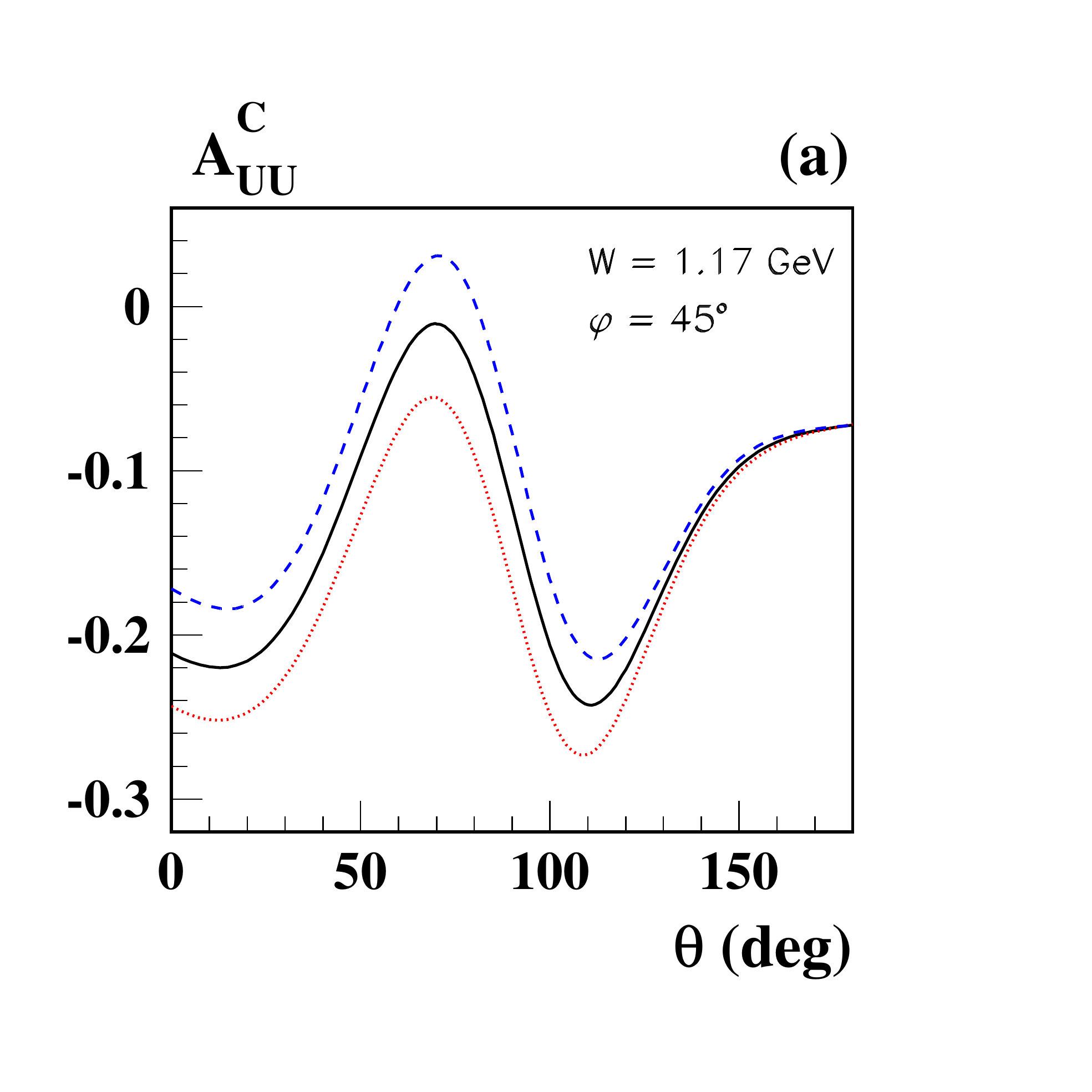}
 \hspace{-20mm}
 \includegraphics[width=0.40\textwidth]{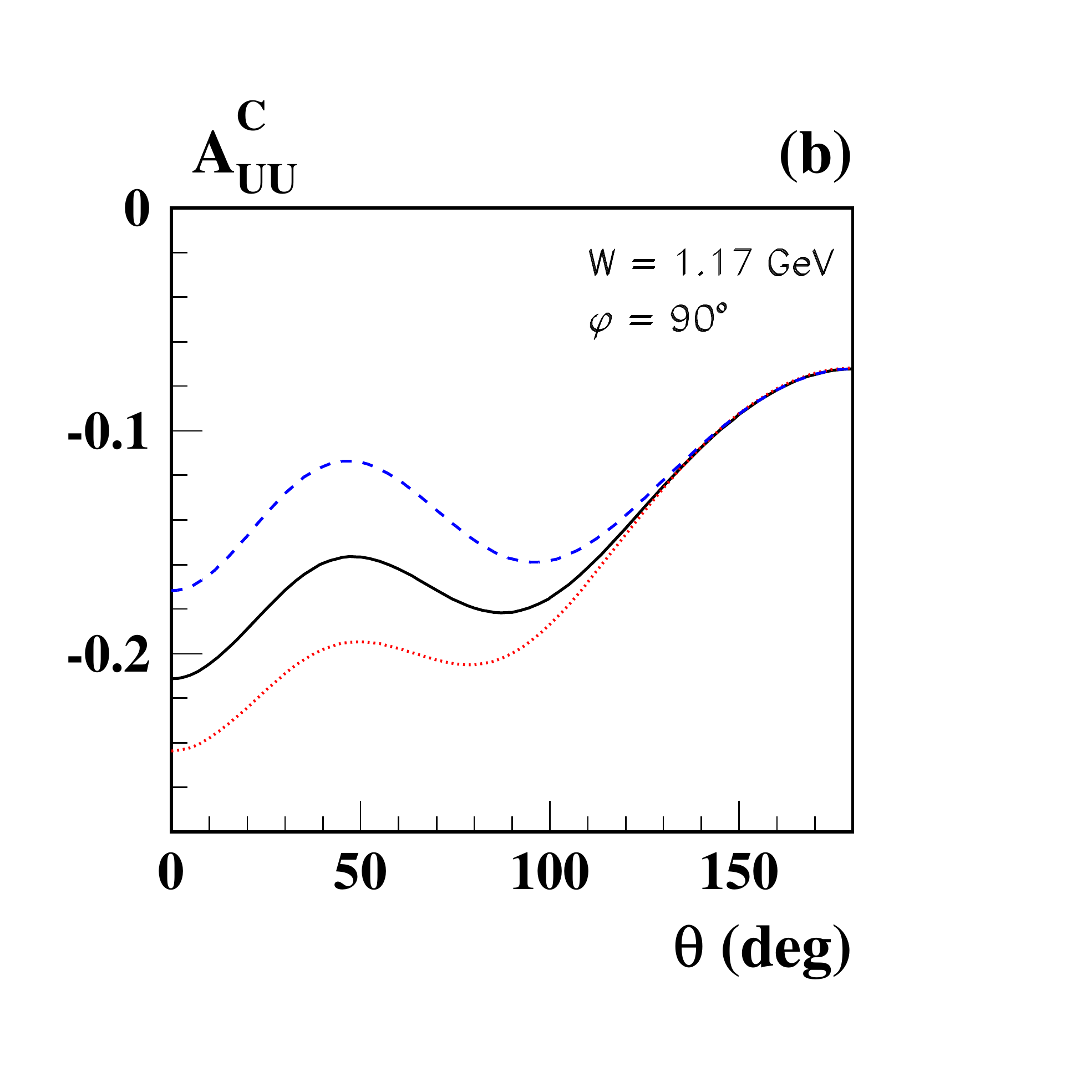}
 \hspace{-20mm}
 \includegraphics[width=0.40\textwidth]{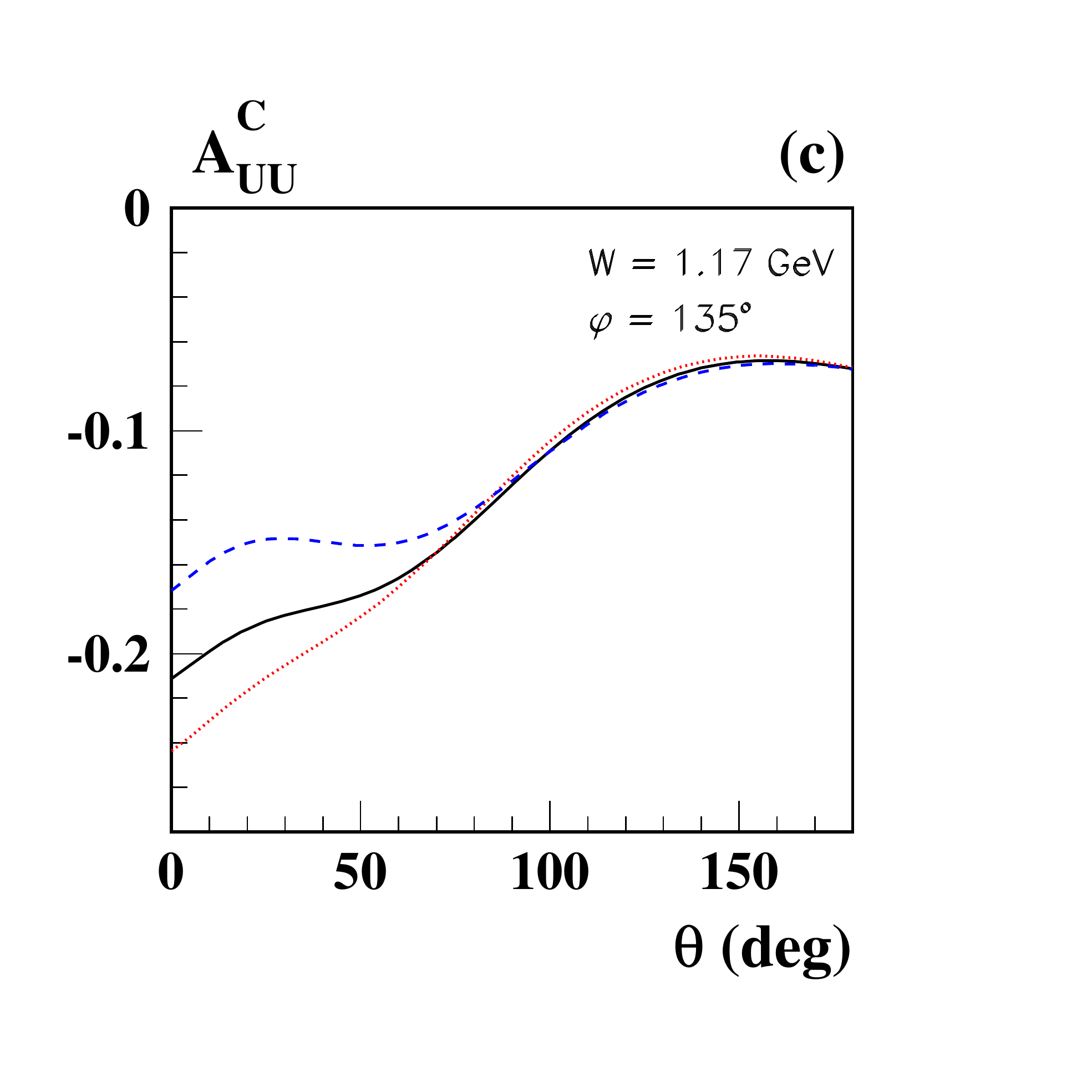}
 \\
\vspace{-8 mm}
 \includegraphics[width=0.40\textwidth]{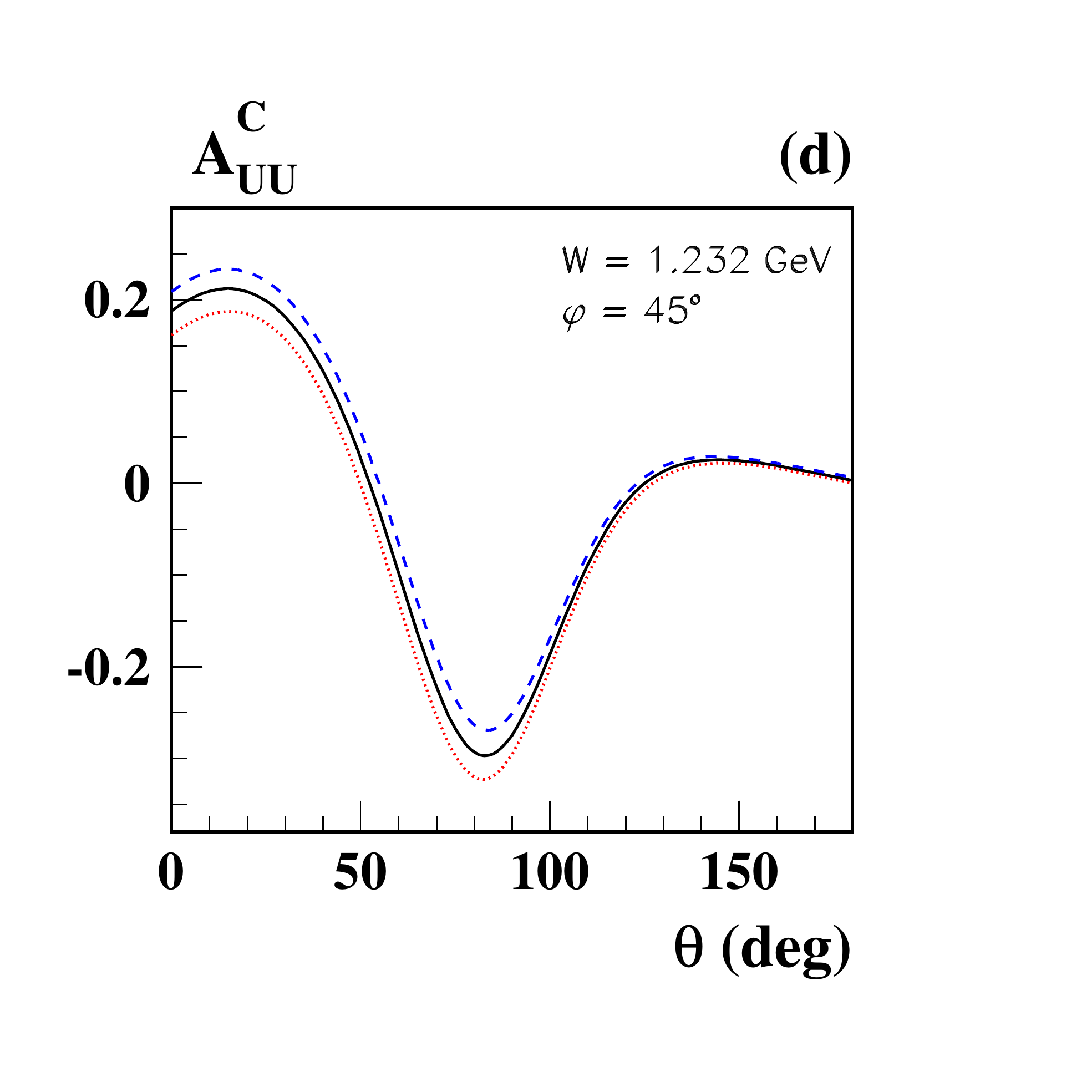}
 \hspace{-20mm}
 \includegraphics[width=0.40\textwidth]{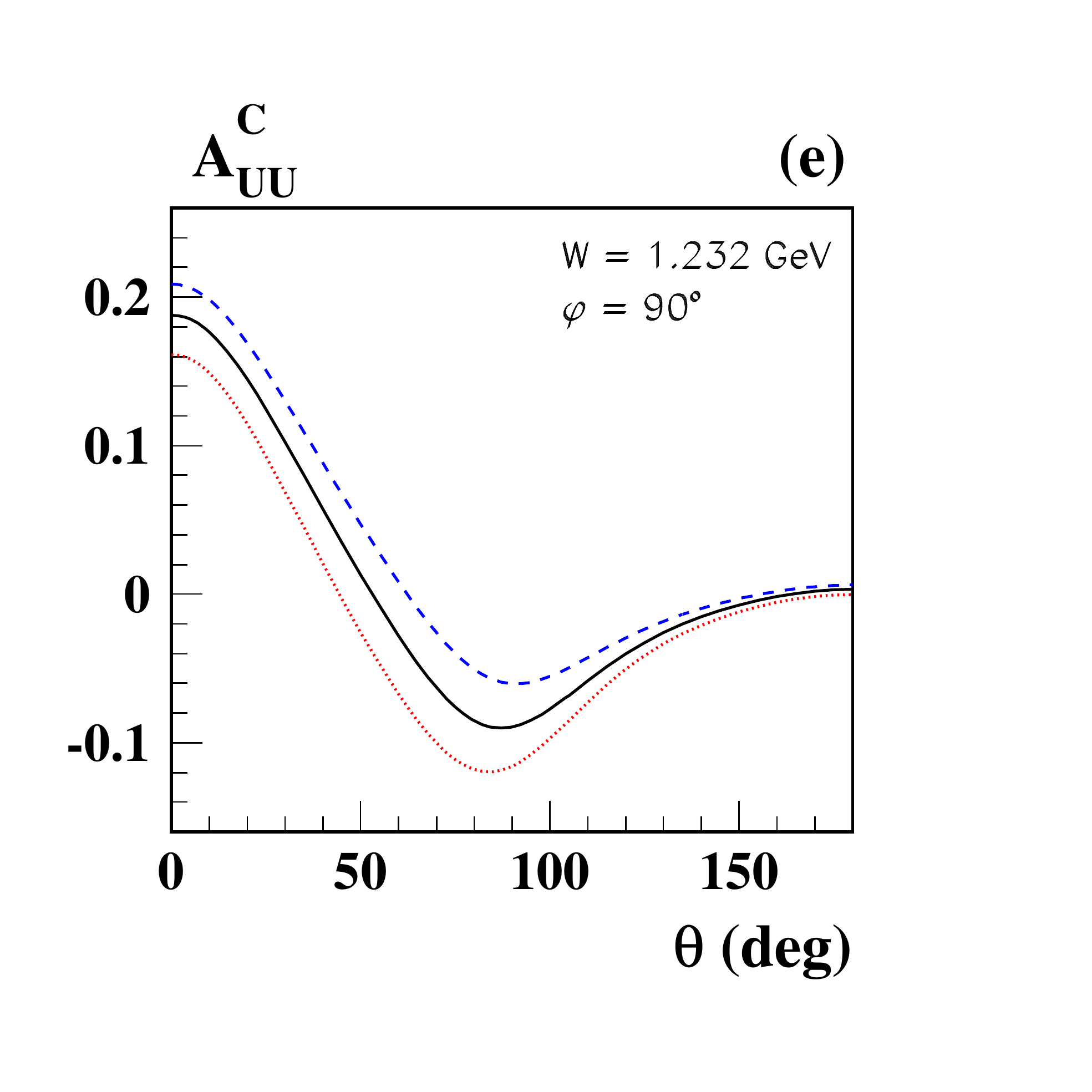}
 \hspace{-20mm}
 \includegraphics[width=0.40\textwidth]{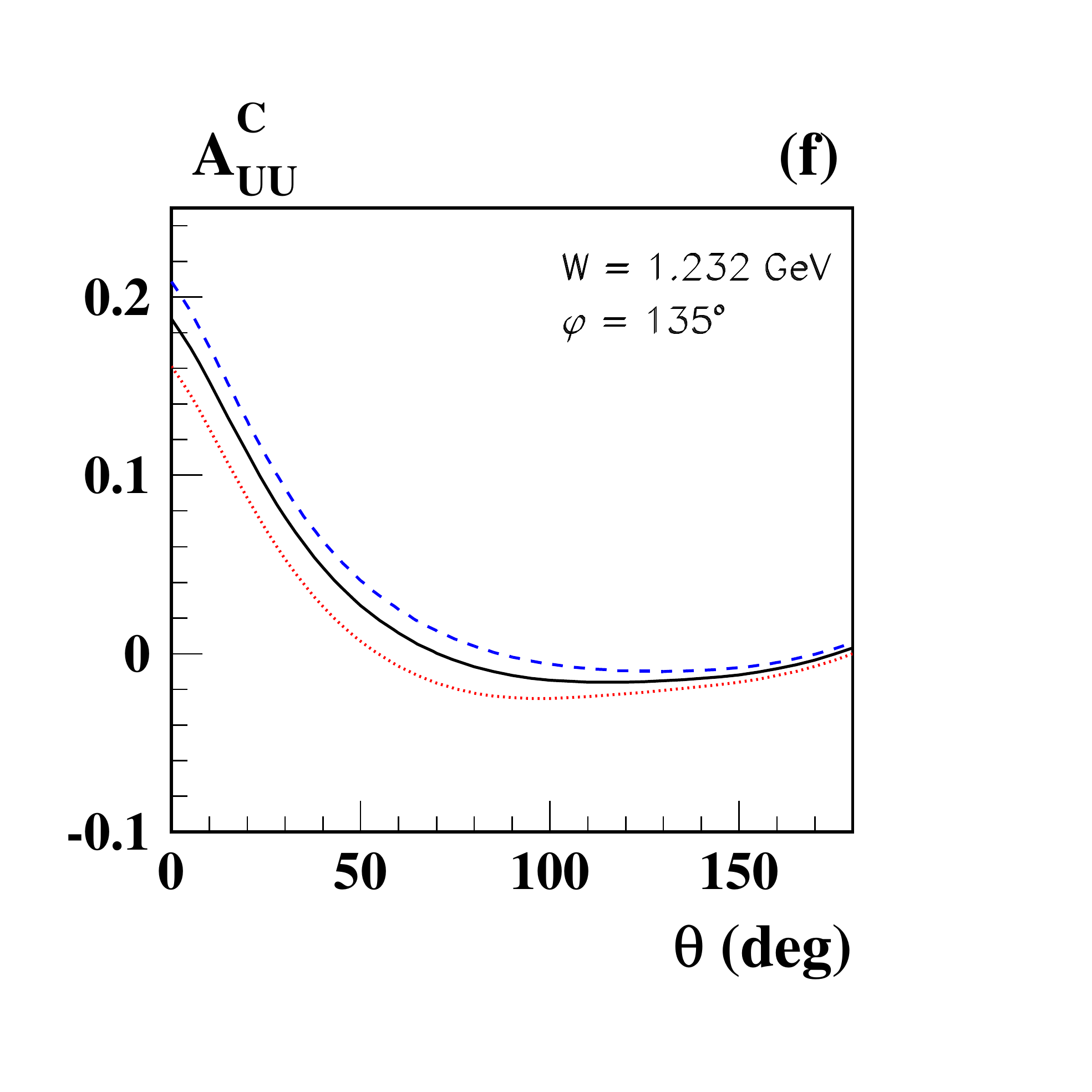}
 \end{center}  
 \caption{The beam-charge asymmetry $A_{UU}^C$ as function of the photon scattering angle for out-of-plane-plane kinematics, for $Q^2=0.43$ GeV$^2$ and   the c.m. energy $W=1.17$ GeV (upper panels) and $W=1.232$ GeV (lower panels), and different values of the azimuthal angle $\phi$: $\phi=45^{{\rm o}}$ (panels (a) and (d)), $\phi=90^{{\rm o}}$ (panels (b) and (e)) and $\phi=135^{{\rm o}}$ (panels (c) and (f)). The results are for $\Lambda_\alpha=\Lambda_\beta=0.5$ GeV (red dotted curves), $\Lambda_\alpha=\Lambda_\beta=0.7$ (black solid curves) and $\Lambda_\alpha=\Lambda_\beta=0.9$ GeV (blue dashed curves). }
   \label{fig:3}  
 \end{figure*}

\begin{figure*}[h]
\begin{center}
 \includegraphics[width=0.4\textwidth]{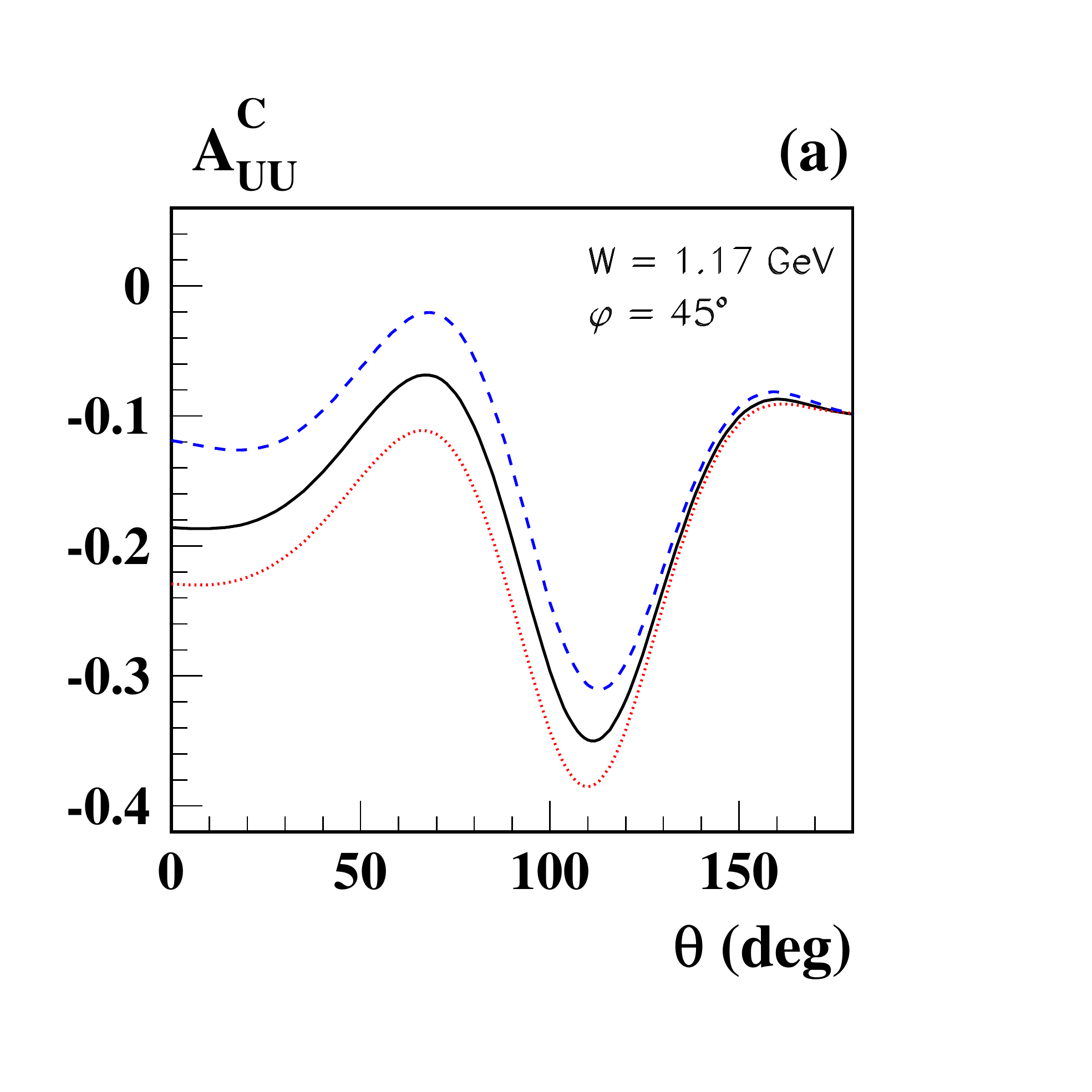}
 \hspace{-20mm}
 \includegraphics[width=0.4\textwidth]{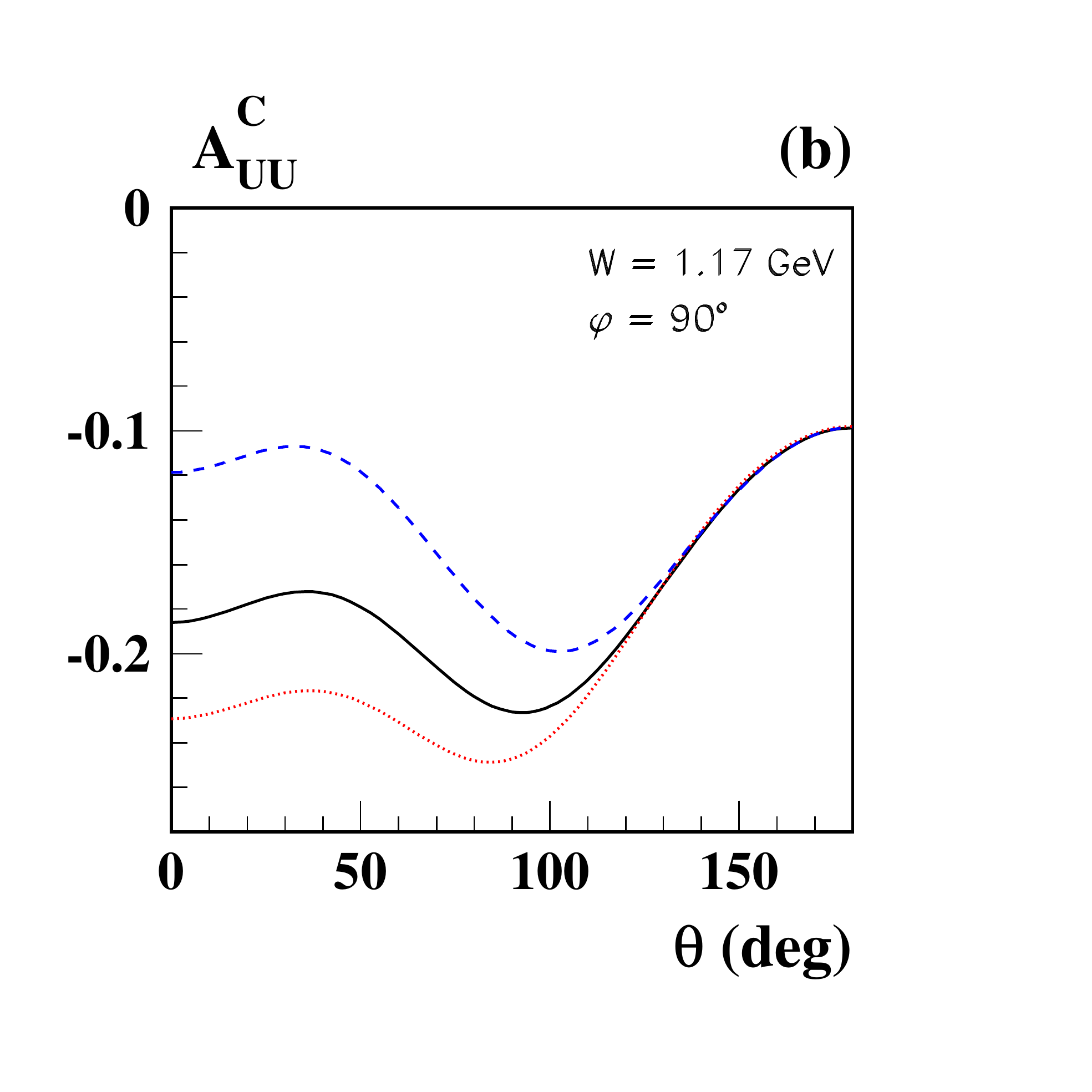}
 \hspace{-20mm}
 \includegraphics[width=0.4\textwidth]{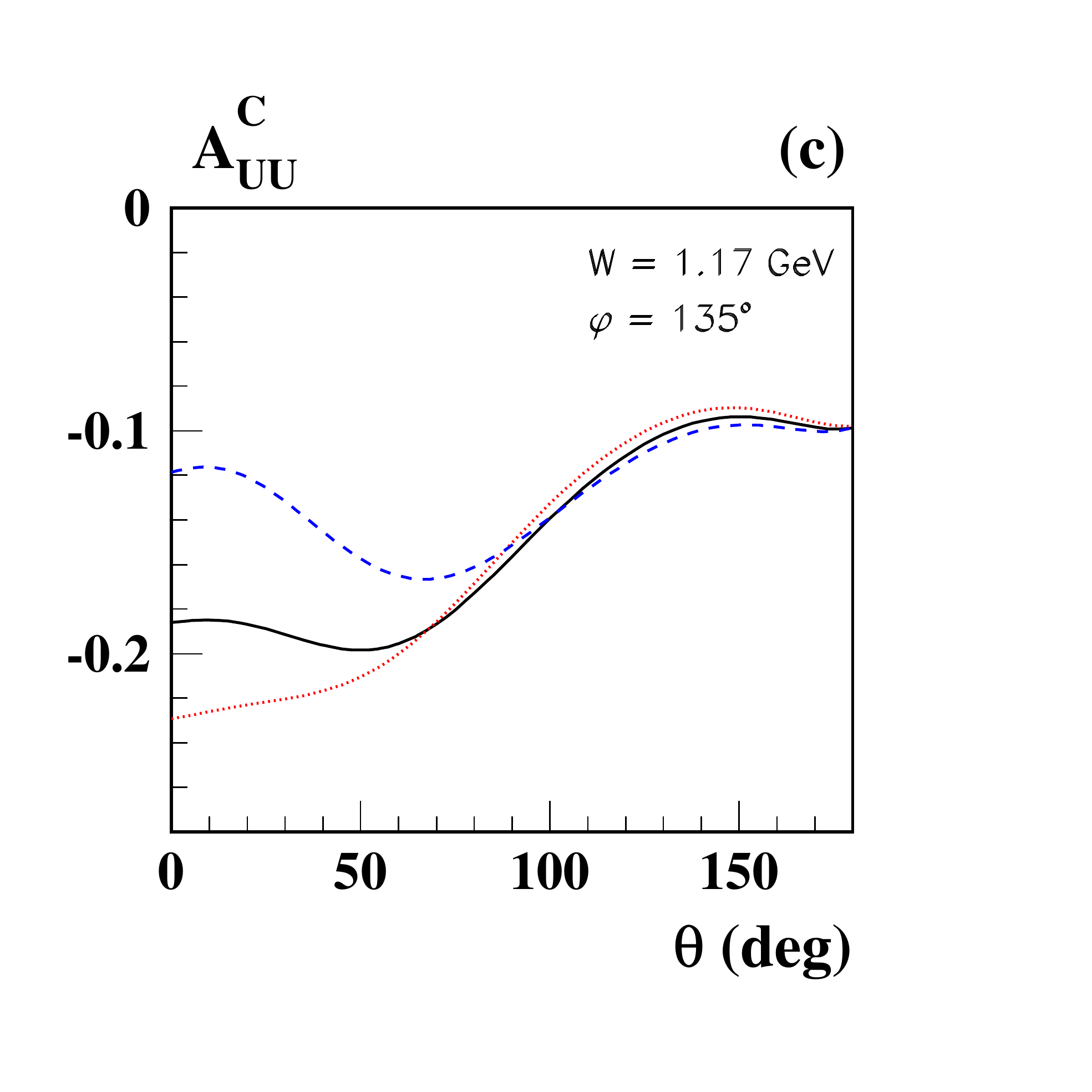}
 \\
\vspace{-8 mm}
 \includegraphics[width=0.4\textwidth]{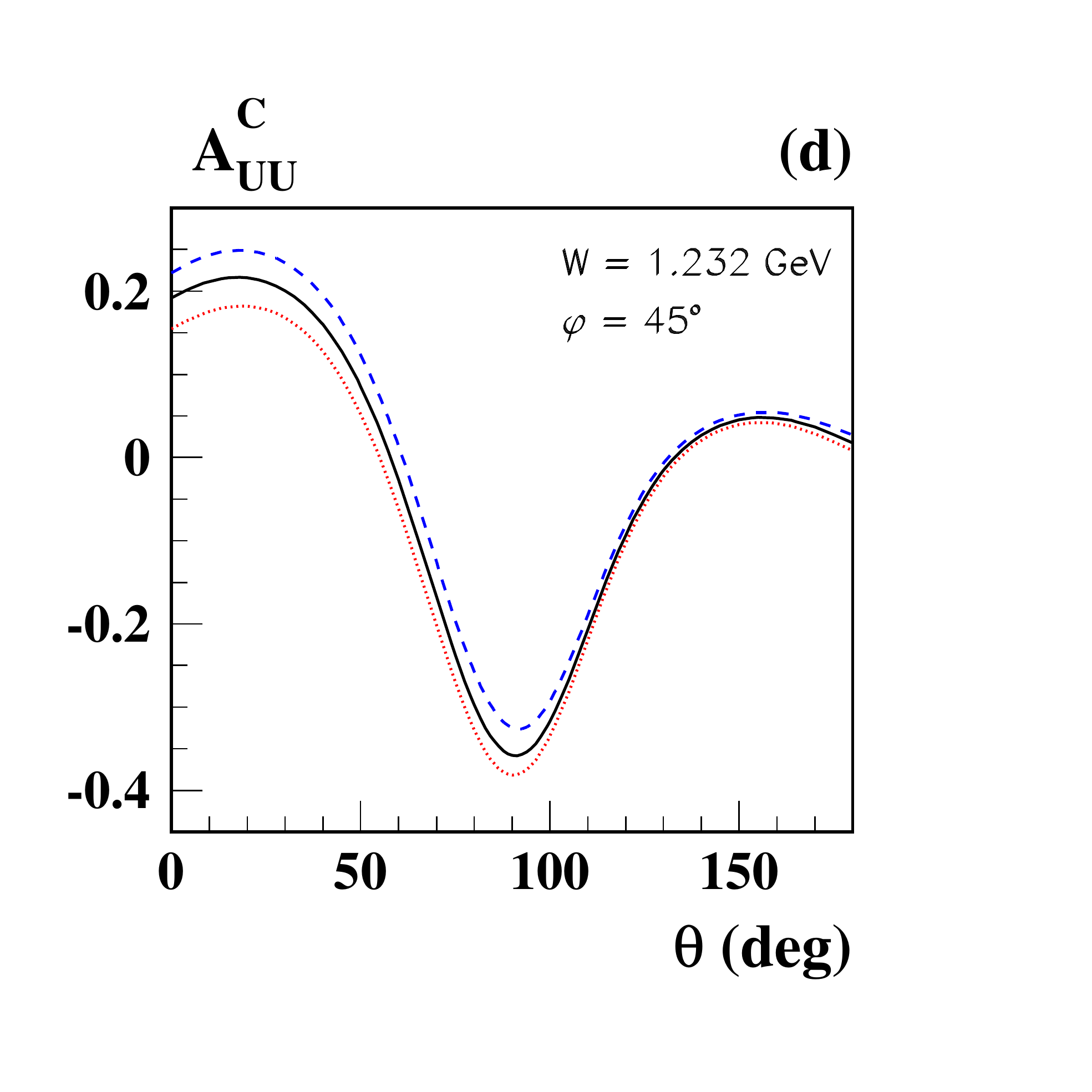}
 \hspace{-20mm}
 \includegraphics[width=0.4\textwidth]{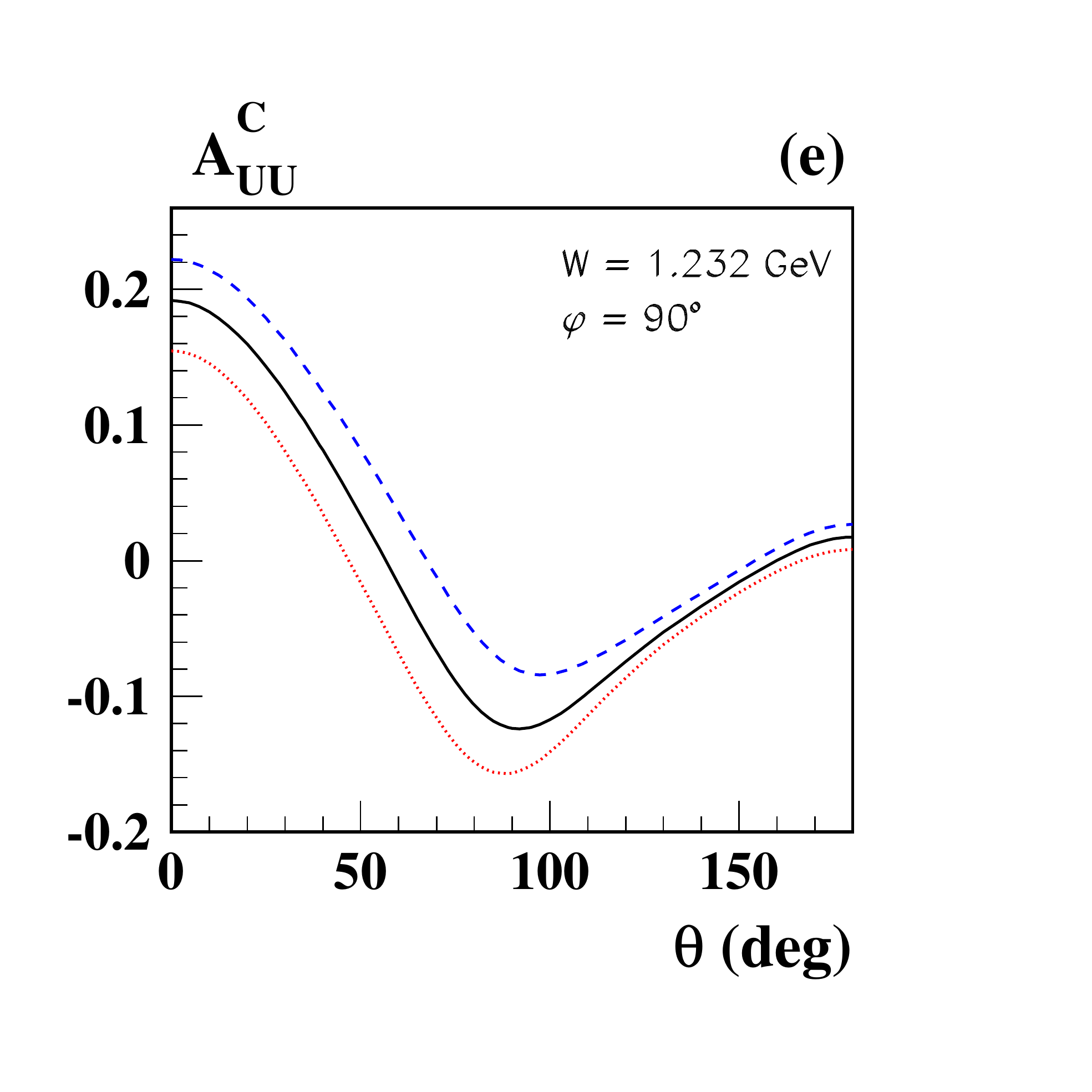}
 \hspace{-20mm}
 \includegraphics[width=0.4\textwidth]{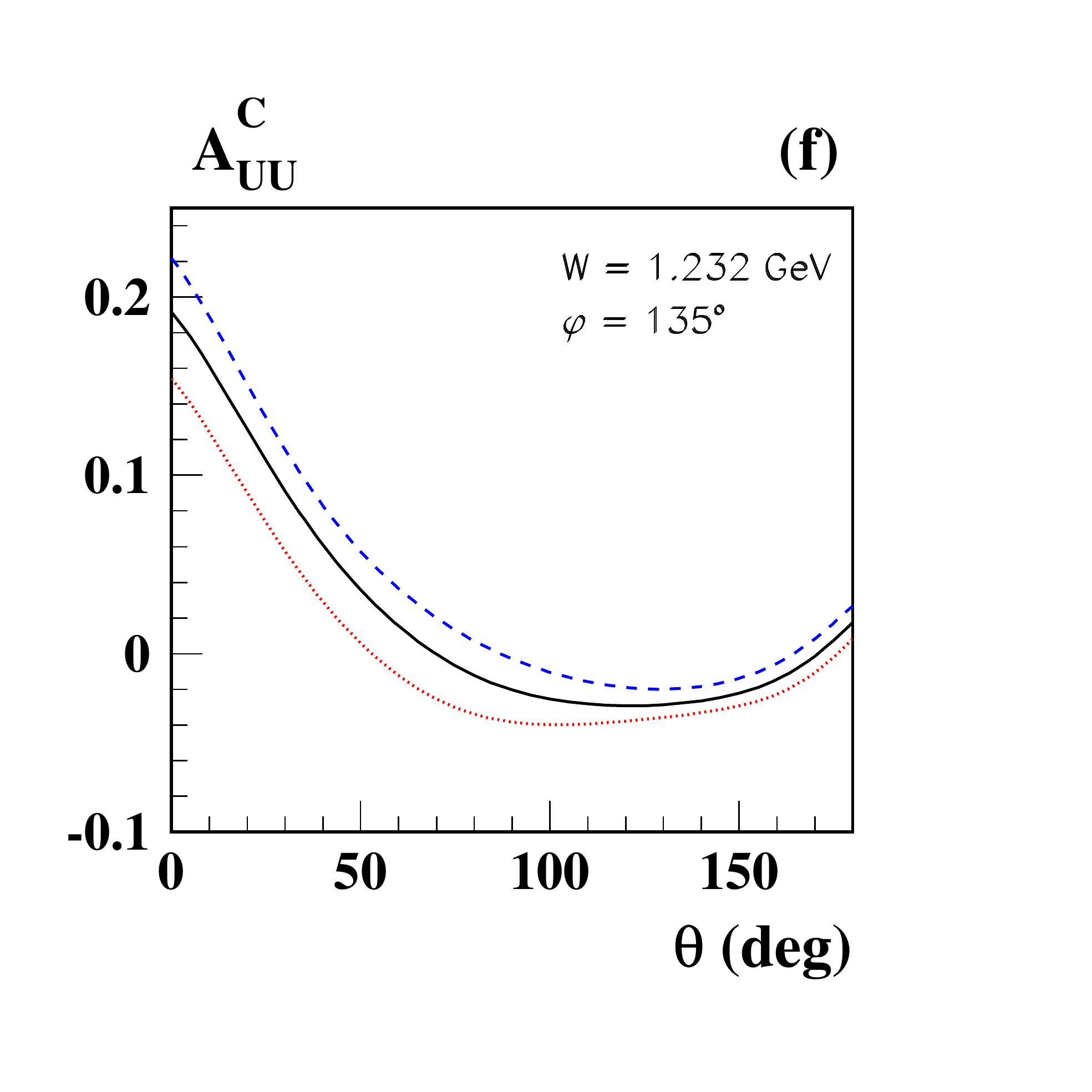}
 \end{center}\caption{The same as in Fig.~\ref{fig:3}, but for $Q^2=0.65$ GeV$^2$. }
  \label{fig:4}  
 \end{figure*}

The same kinematics is explored in Fig.~\ref{fig:3} for $Q^2=0.43$ GeV$^2$ and in Fig.~\ref{fig:4} for $Q^2=0.65$ GeV$^2$, but for out-of-plane angles $\phi$. In these cases, the sensitivity to the scalar polarizabilities is enhanced, especially at the lower center of mass energy and higher values of momentum transfer. The largest effects are  for $\phi=90^o$ at  scattering angles $\theta \le 100^o$, where we observe a relative effect on the asymmetry of up to $\pm 30\%$ by scaling the mass parameters by $\pm 0.2$ units.

\subsection{Beam-spin asymmetry}
 
\begin{figure*}[h!]
\begin{center}
 \includegraphics[width=0.3\textwidth]{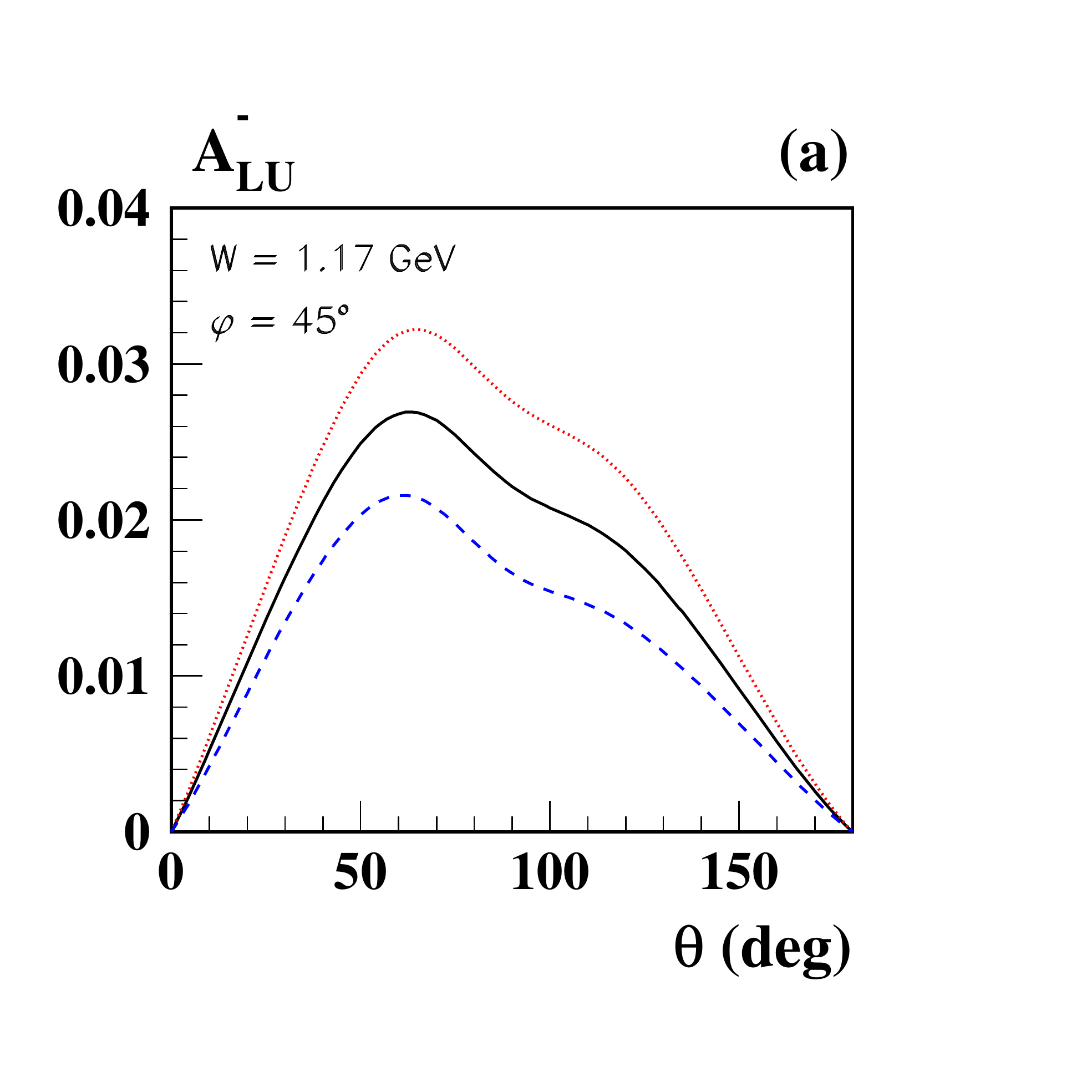}
 \hspace{-14mm}
 \includegraphics[width=0.3\textwidth]{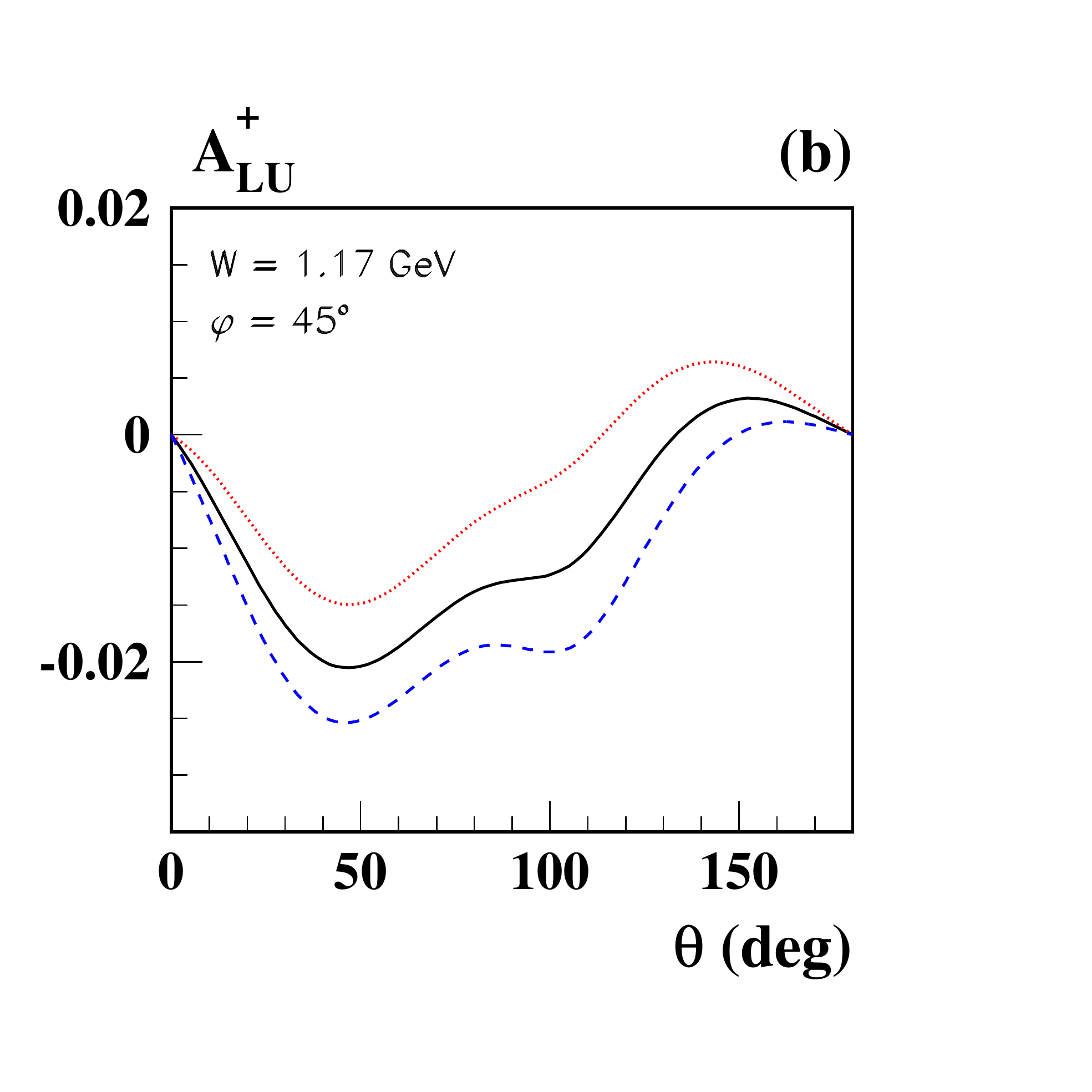}
  \hspace{-14mm}
 \includegraphics[width=0.3\textwidth]{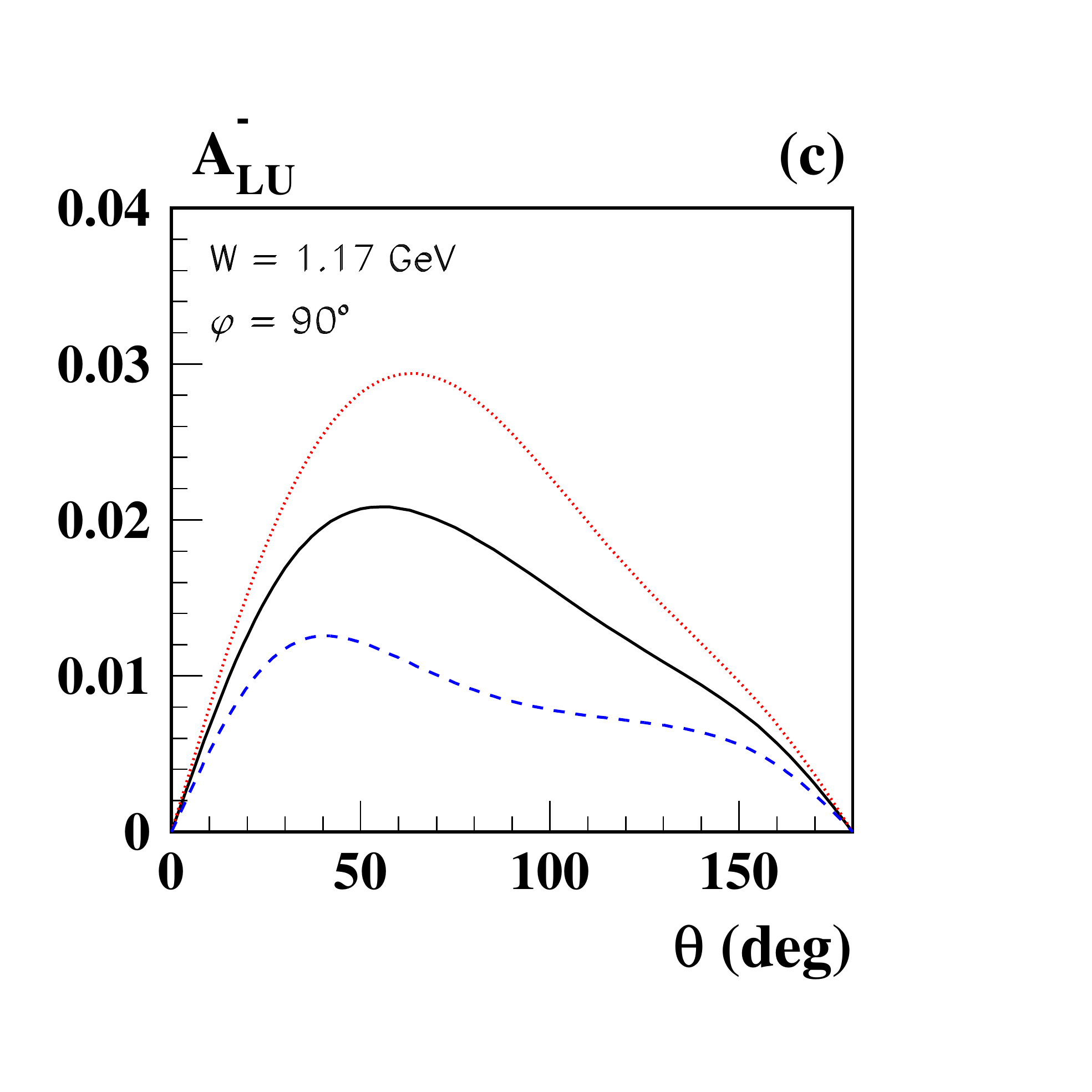}
 \hspace{-14mm}
 \includegraphics[width=0.3\textwidth]{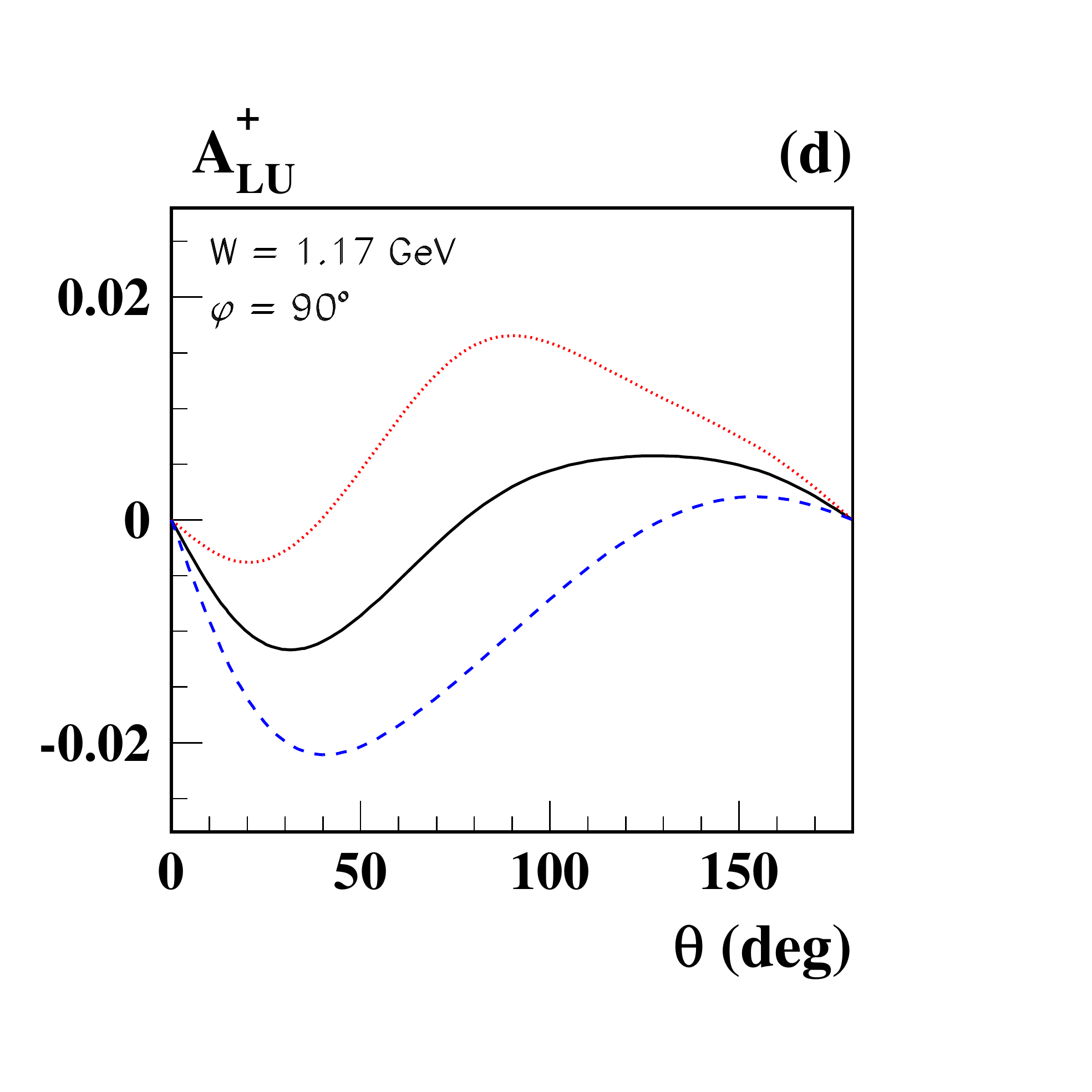}
 \\
 \vspace{-0.5 truecm}
 \includegraphics[width=0.3\textwidth]{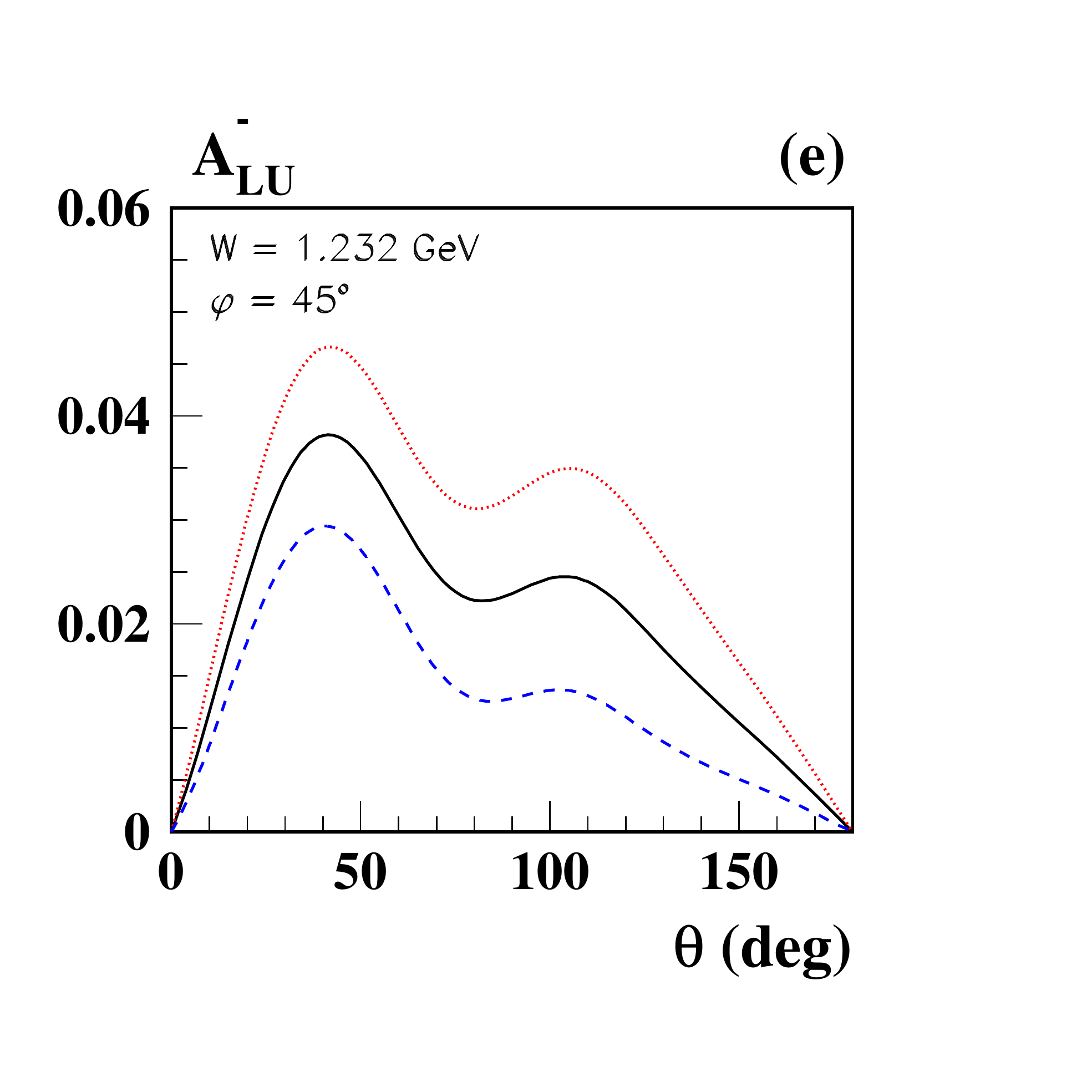}
 \hspace{-14mm}
 \includegraphics[width=0.3\textwidth]{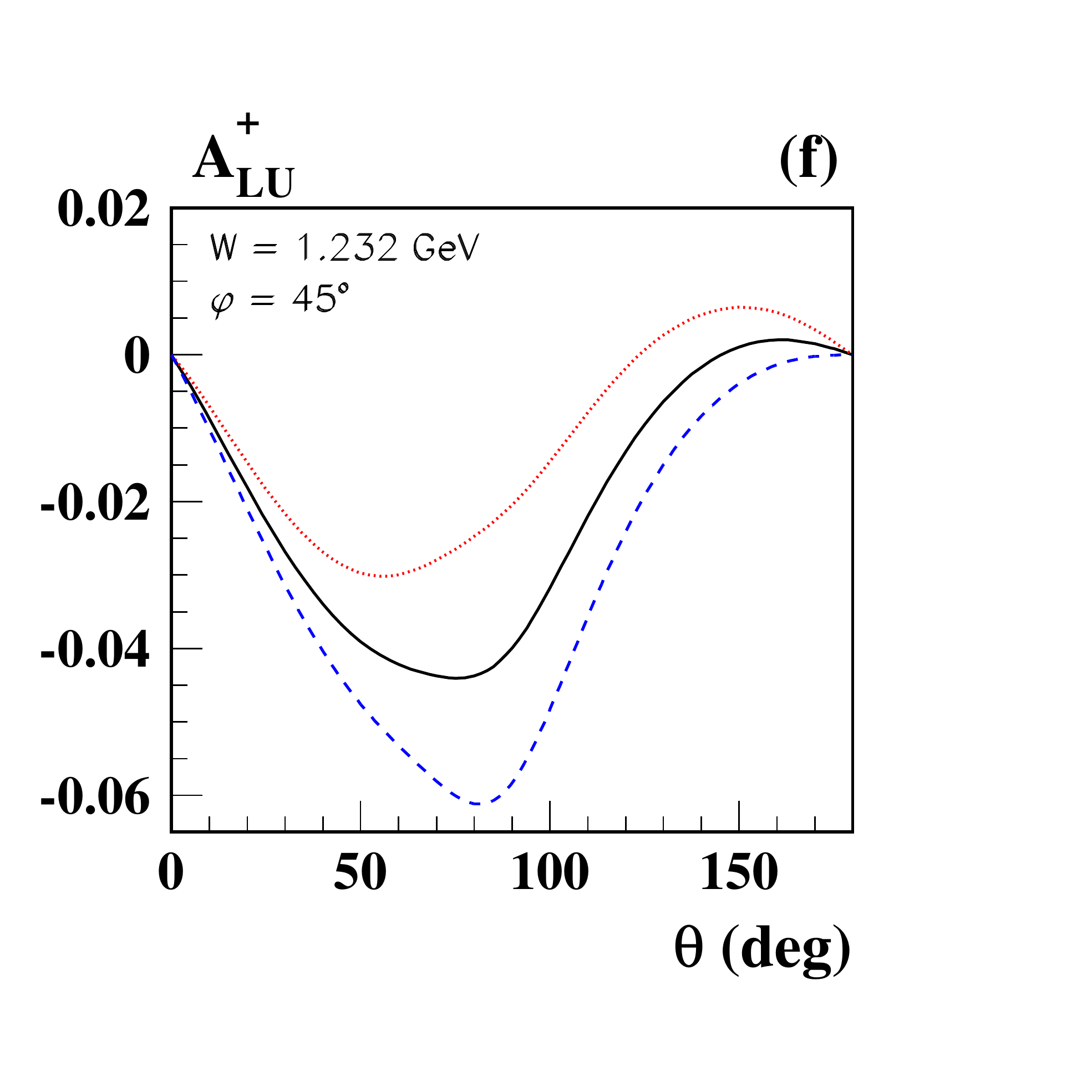}
  \hspace{-14mm}
 \includegraphics[width=0.3\textwidth]{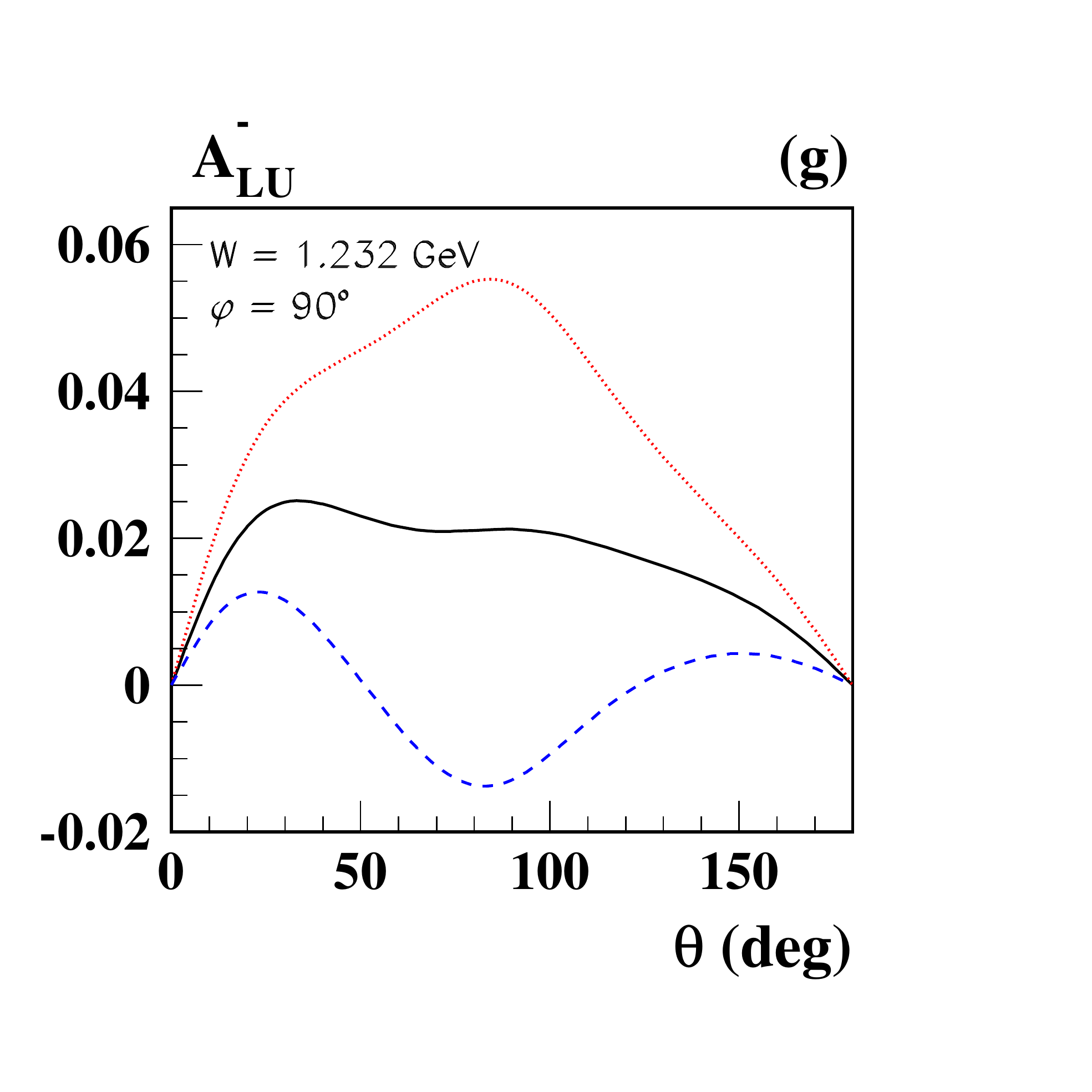}
 \hspace{-14mm}
 \includegraphics[width=0.3\textwidth]{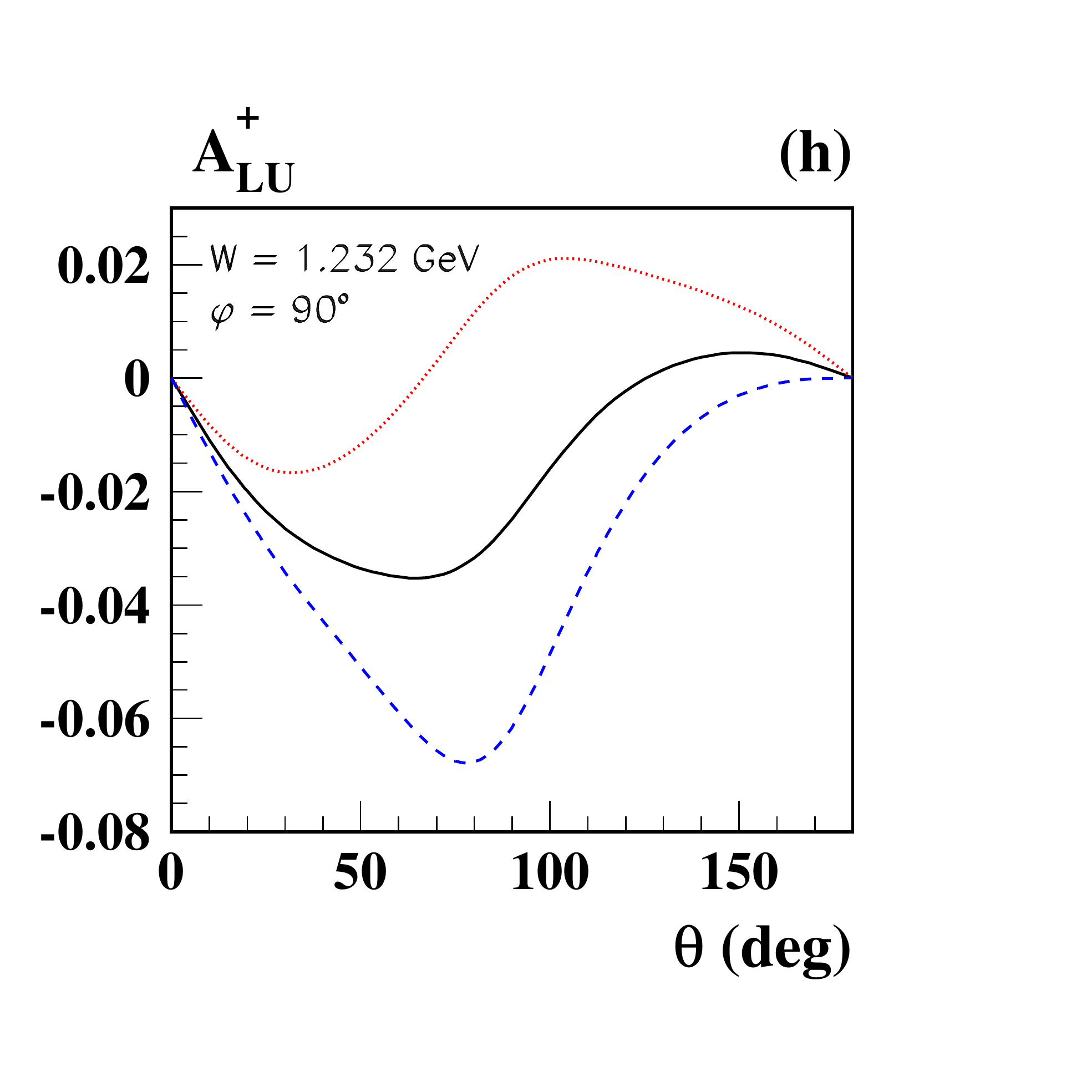}
\caption{The electron beam-spin asymmetry $A_{LU}^-$ and the positron beam-spin asymmetry $A^+_{LU}$ as function of the photon scattering angle for out-of-plane-plane kinematics, for $Q^2=0.43$ GeV$^2$ and c.m. energy $W=1.17$ GeV (upper panels) and $W=1.232$ GeV (lower panels), at different values of the azimuthal angle $\phi$: $\phi=45^{{\rm o}}$ (panels (a) and (b)), and $\phi=90^{{\rm o}}$ (panels (c) and (c)) .The results are for $\Lambda_\alpha=\Lambda_\beta=0.5$ GeV (red dotted curves), $\Lambda_\alpha=\Lambda_\beta=0.7$ (black solid curves) and $\Lambda_\alpha=\Lambda_\beta=0.9$ GeV (blue dashed curves). }
\label{fig:5}
 \end{center}
 \end{figure*}

\begin{figure*}[h!]
\begin{center}
 \includegraphics[width=0.3\textwidth]{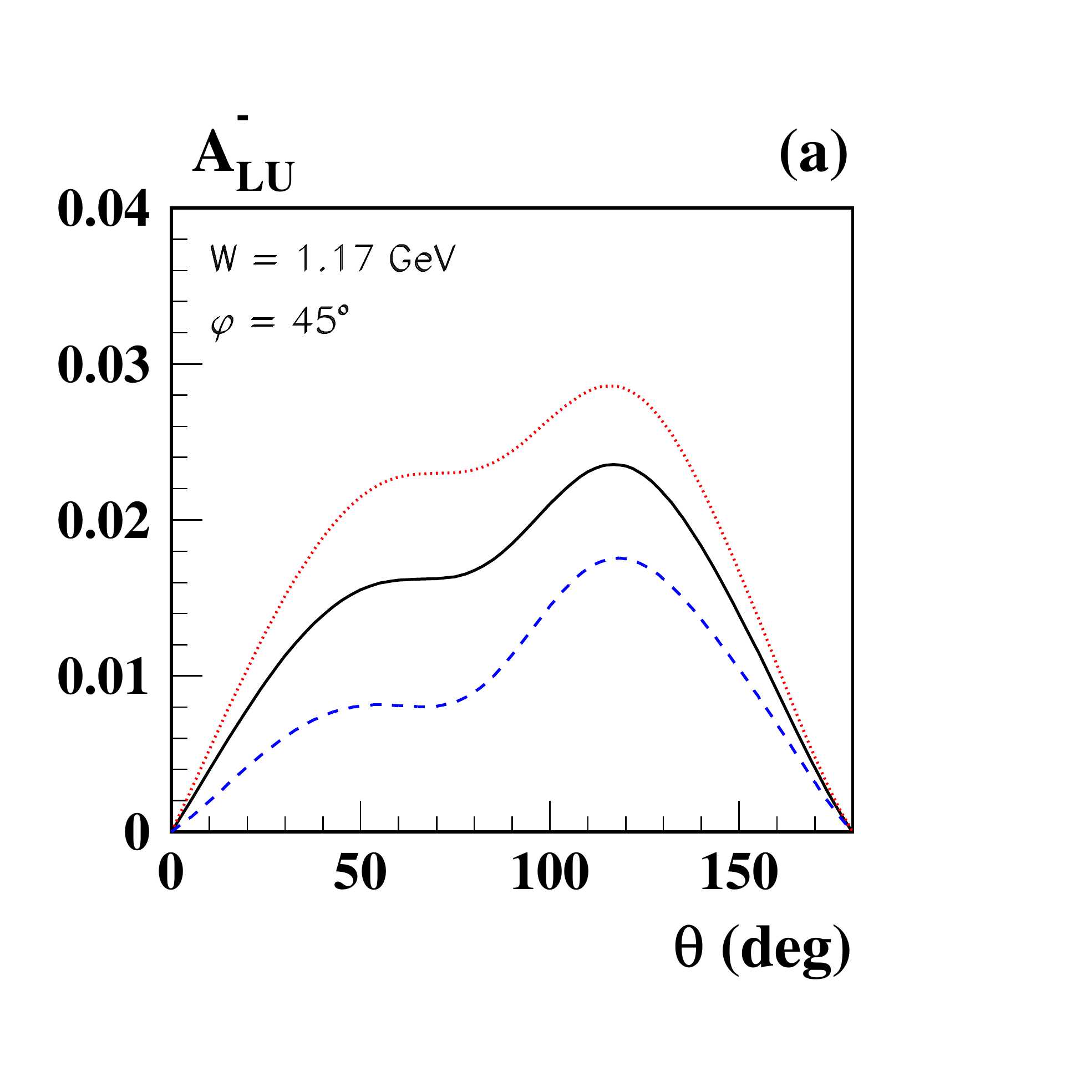}
 \hspace{-14mm}
 \includegraphics[width=0.3\textwidth]{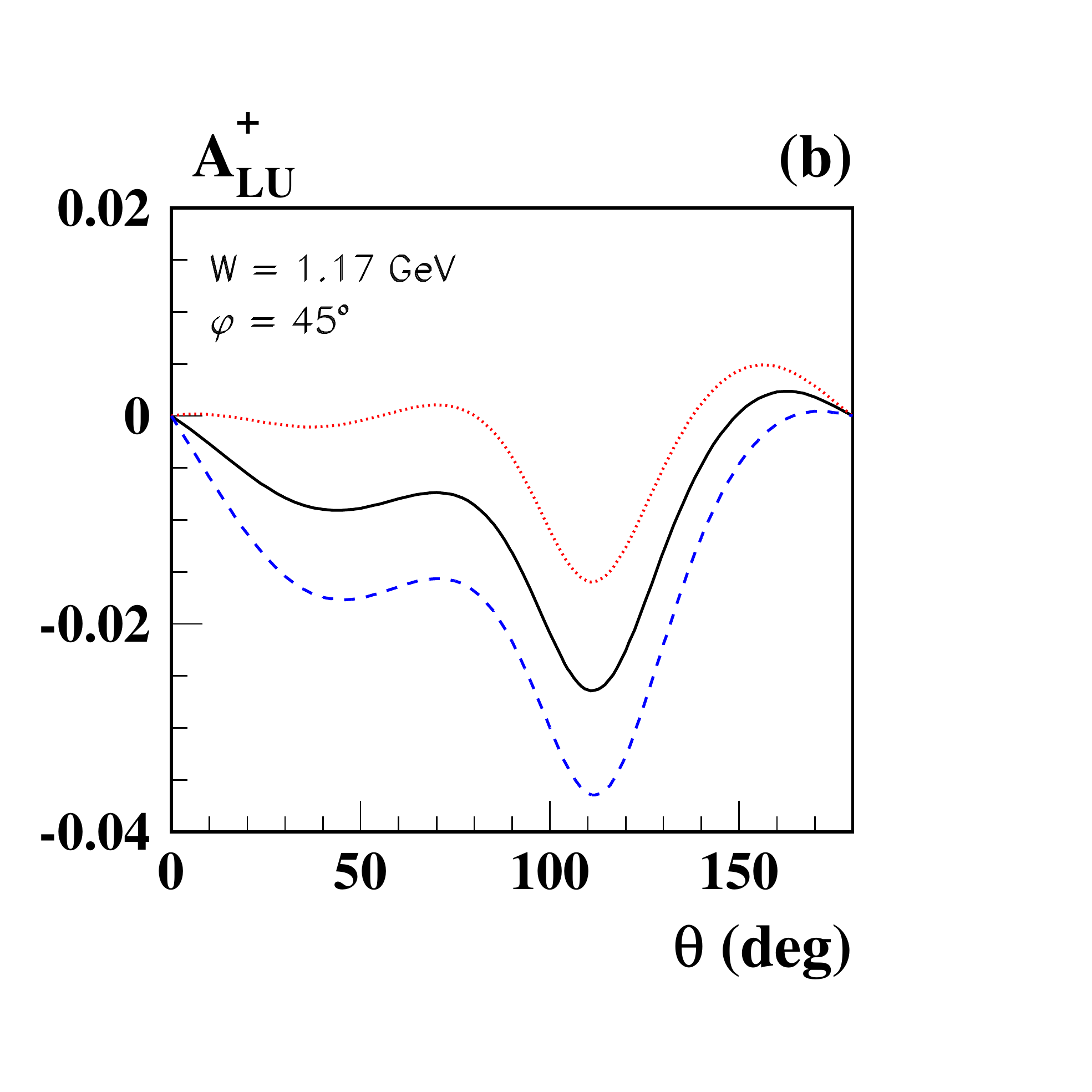}
  \hspace{-14mm}
 \includegraphics[width=0.3\textwidth]{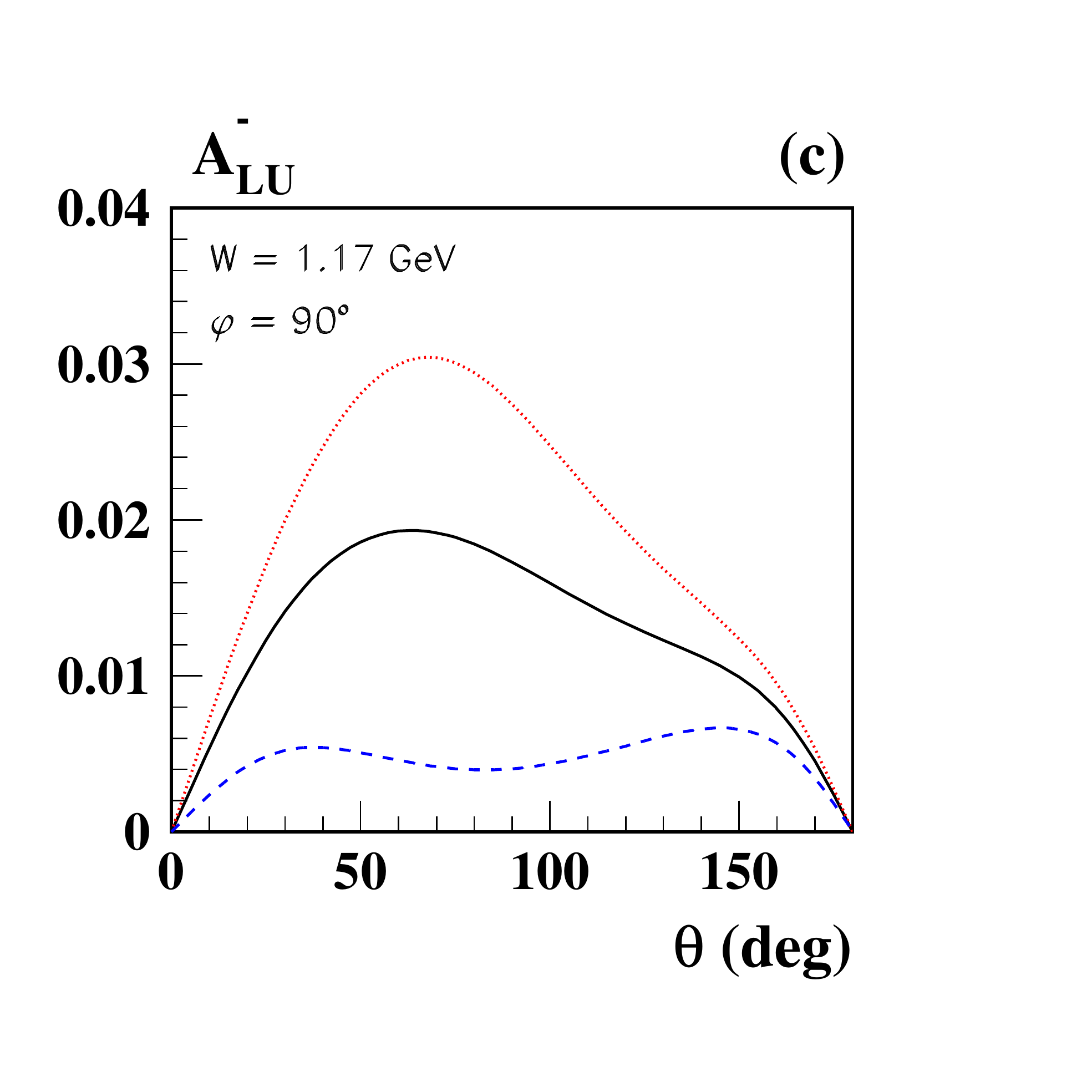}
 \hspace{-14mm}
 \includegraphics[width=0.3\textwidth]{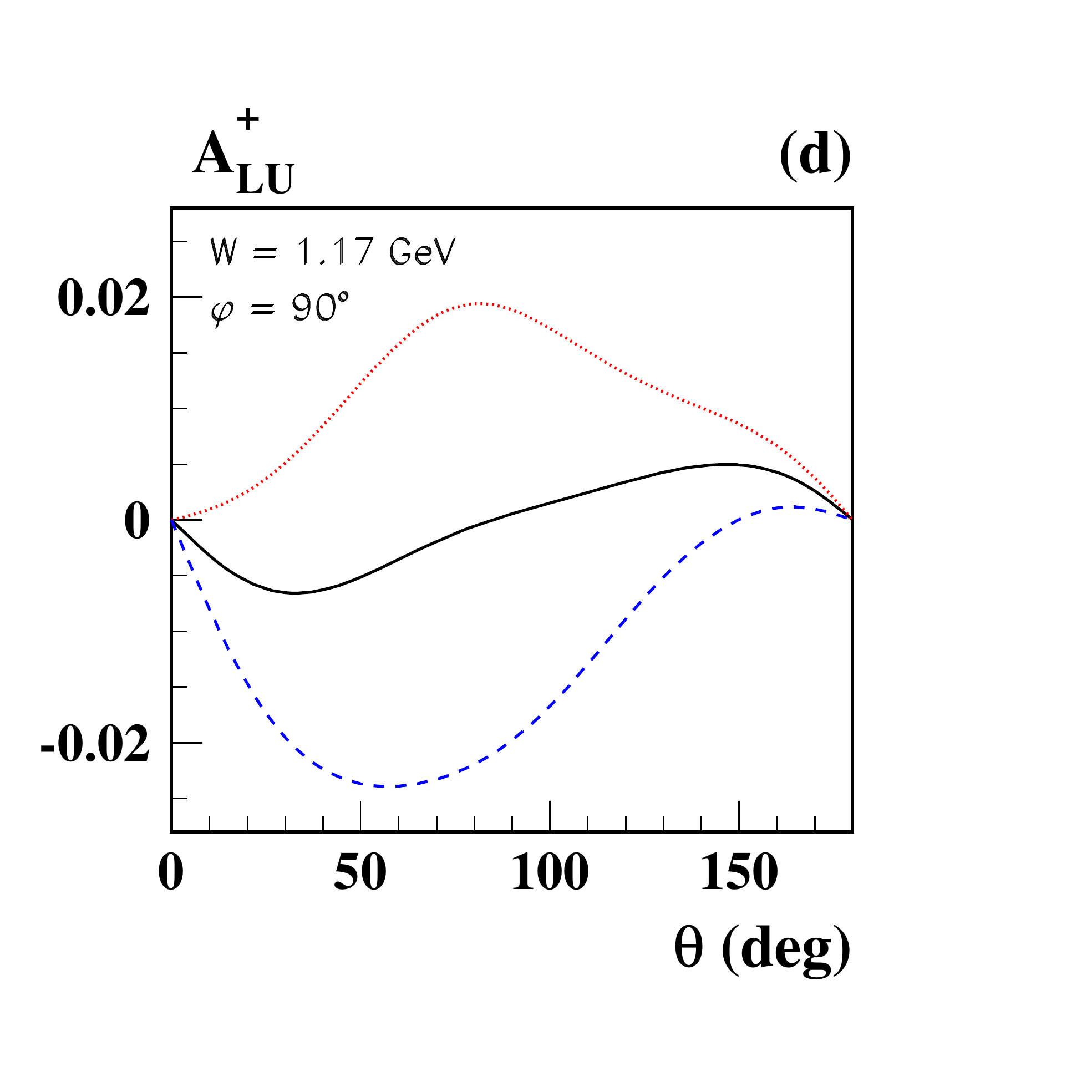}
 \\
 \vspace{-0.5 truecm}
 \includegraphics[width=0.3\textwidth]{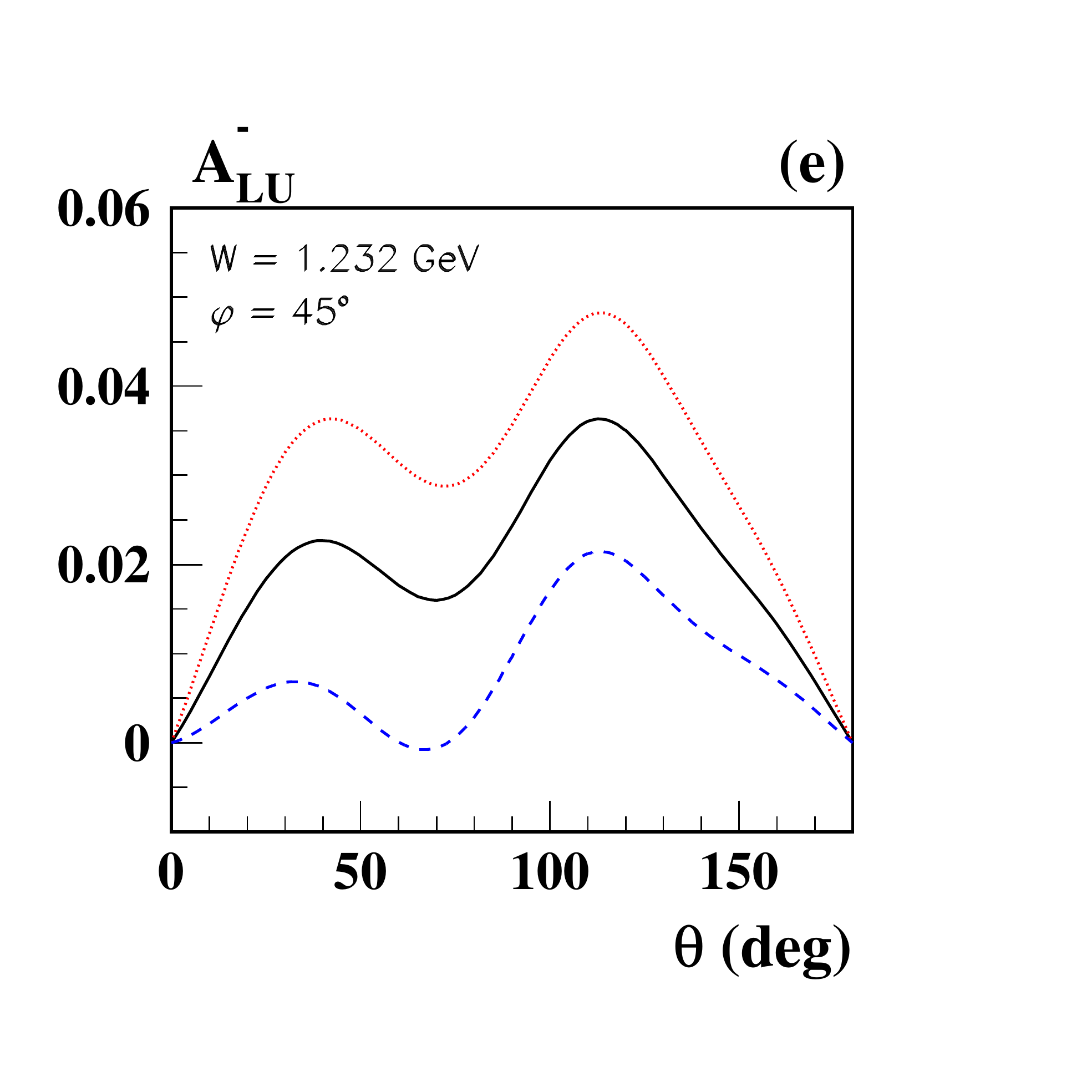}
 \hspace{-14mm}
 \includegraphics[width=0.3\textwidth]{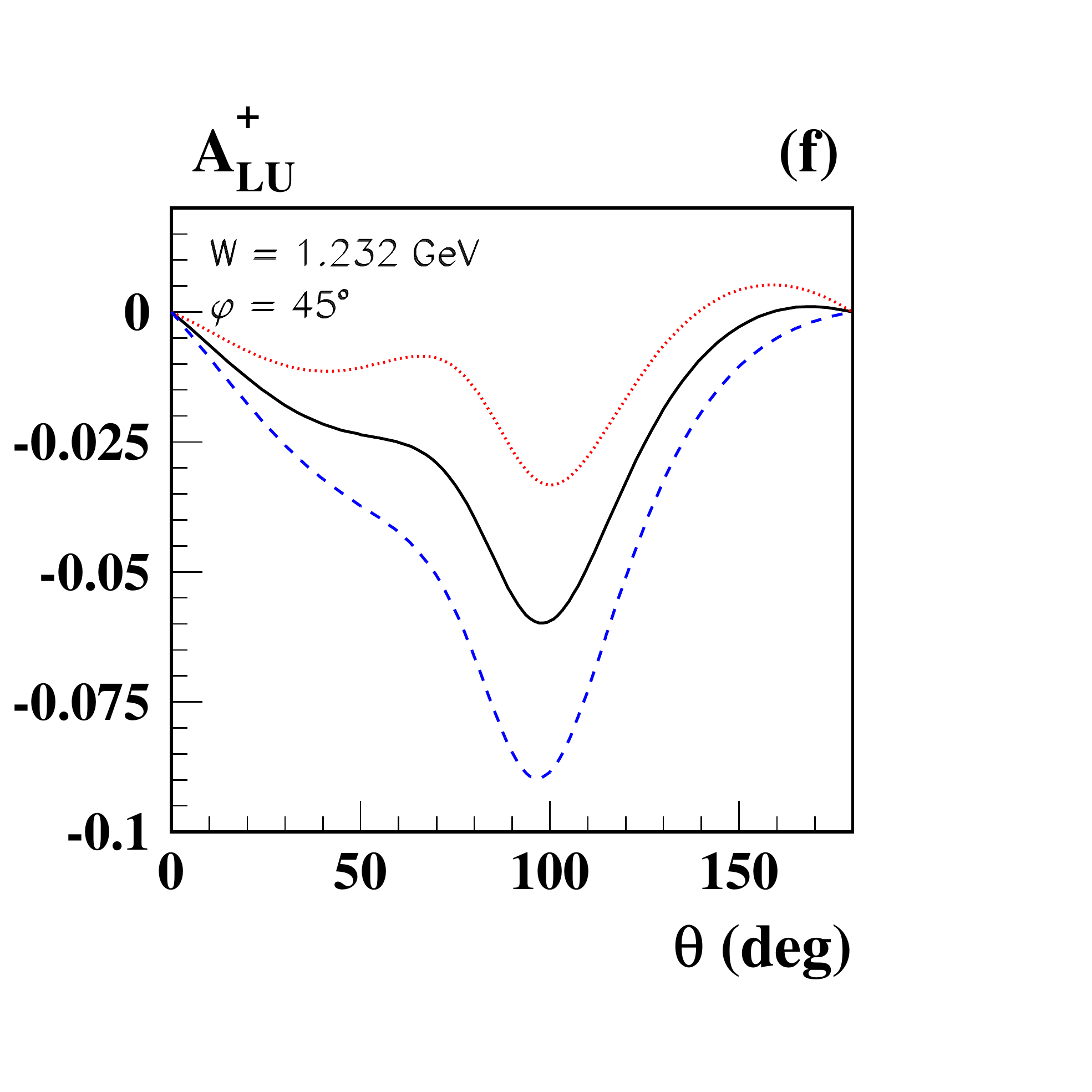}
  \hspace{-14mm}
 \includegraphics[width=0.3\textwidth]{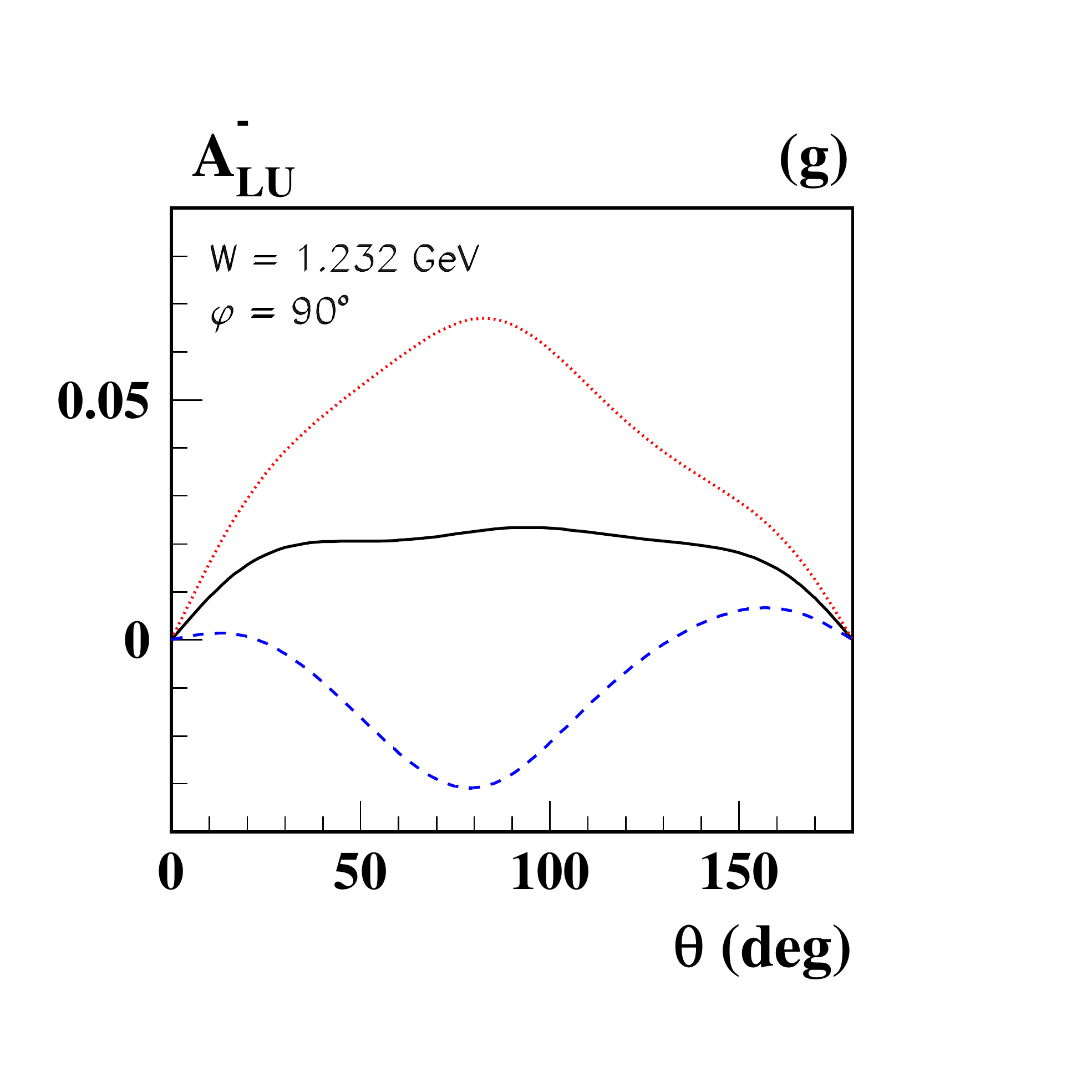}
 \hspace{-14mm}
 \includegraphics[width=0.3\textwidth]{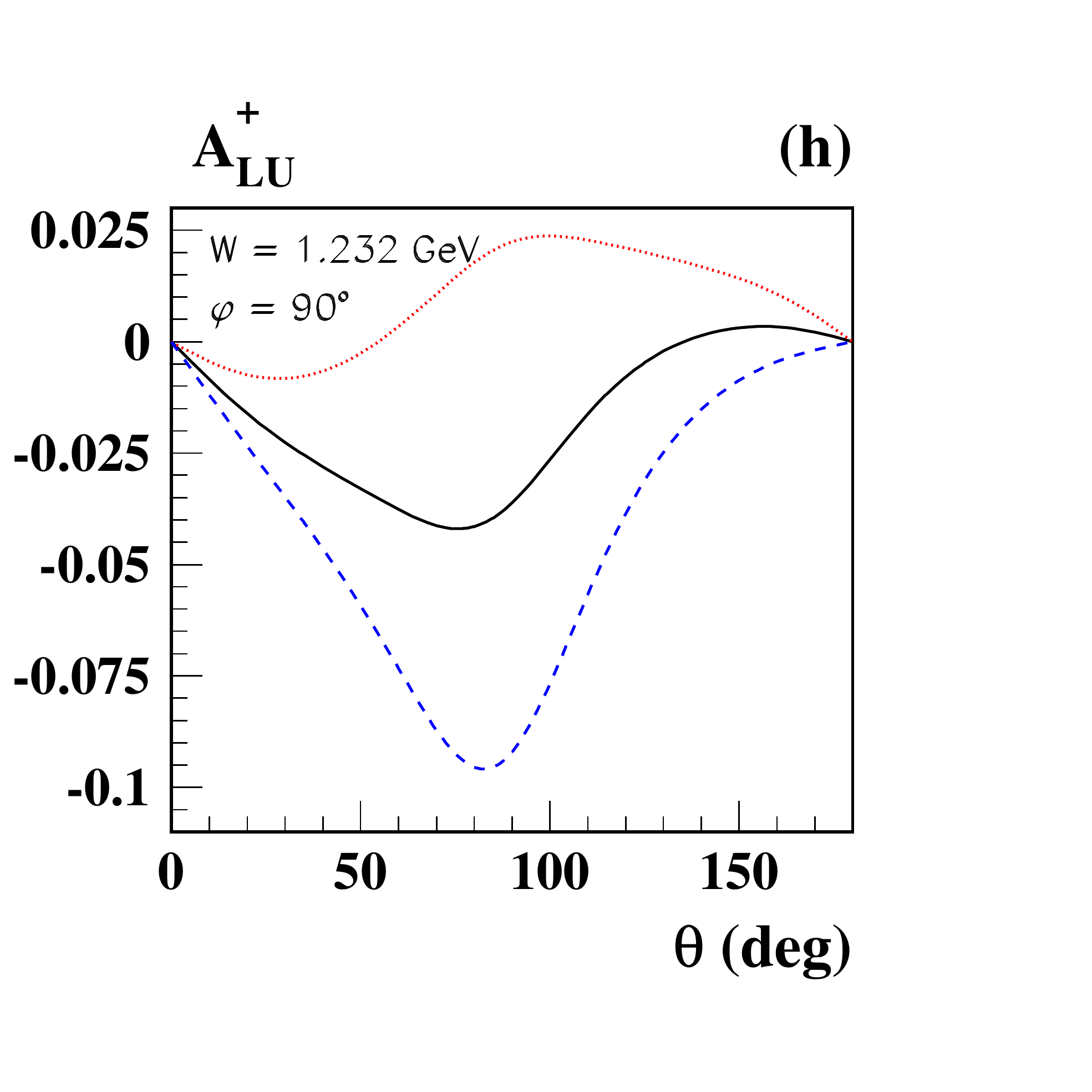}
 \end{center}
    \caption{The same as in Fig.~\ref{fig:5}, but for $Q^2=0.65$ GeV$^2$. }
\label{fig:6}       
\end{figure*}

\begin{figure*}[h!]
\begin{center}
 \includegraphics[width=0.42\textwidth]{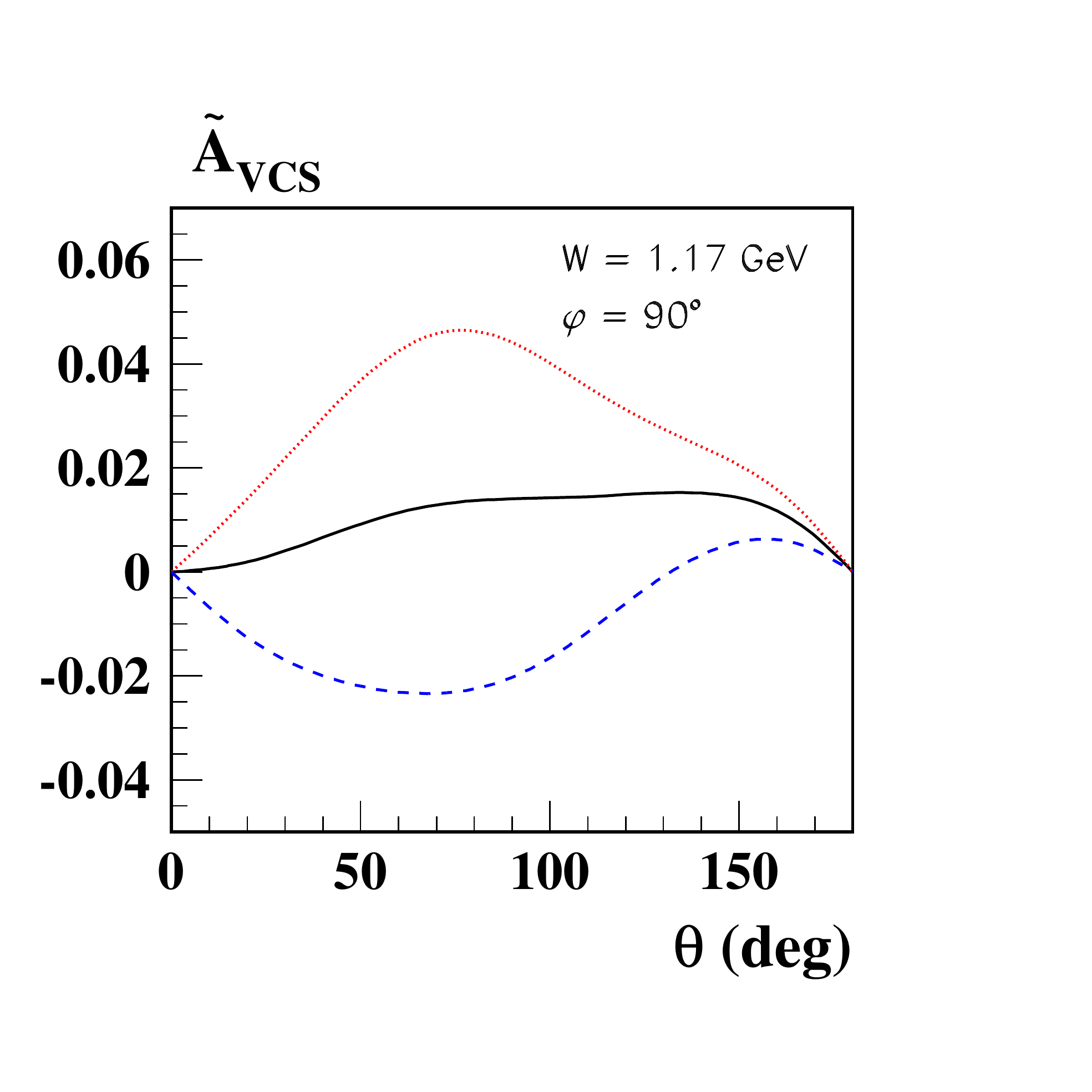}
 \includegraphics[width=0.42\textwidth]{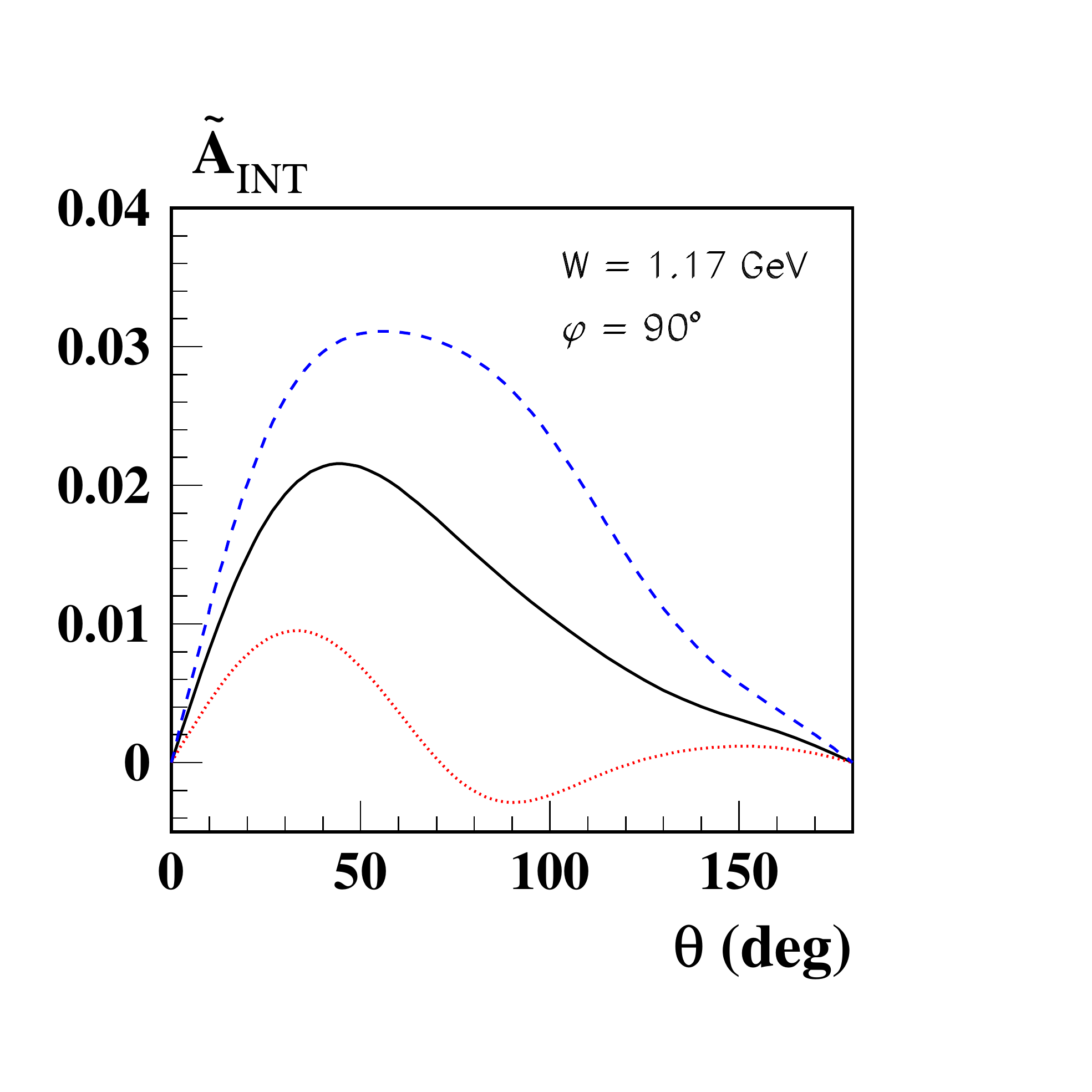}
\\
 \vspace{-15mm}
 \includegraphics[width=0.42\textwidth]{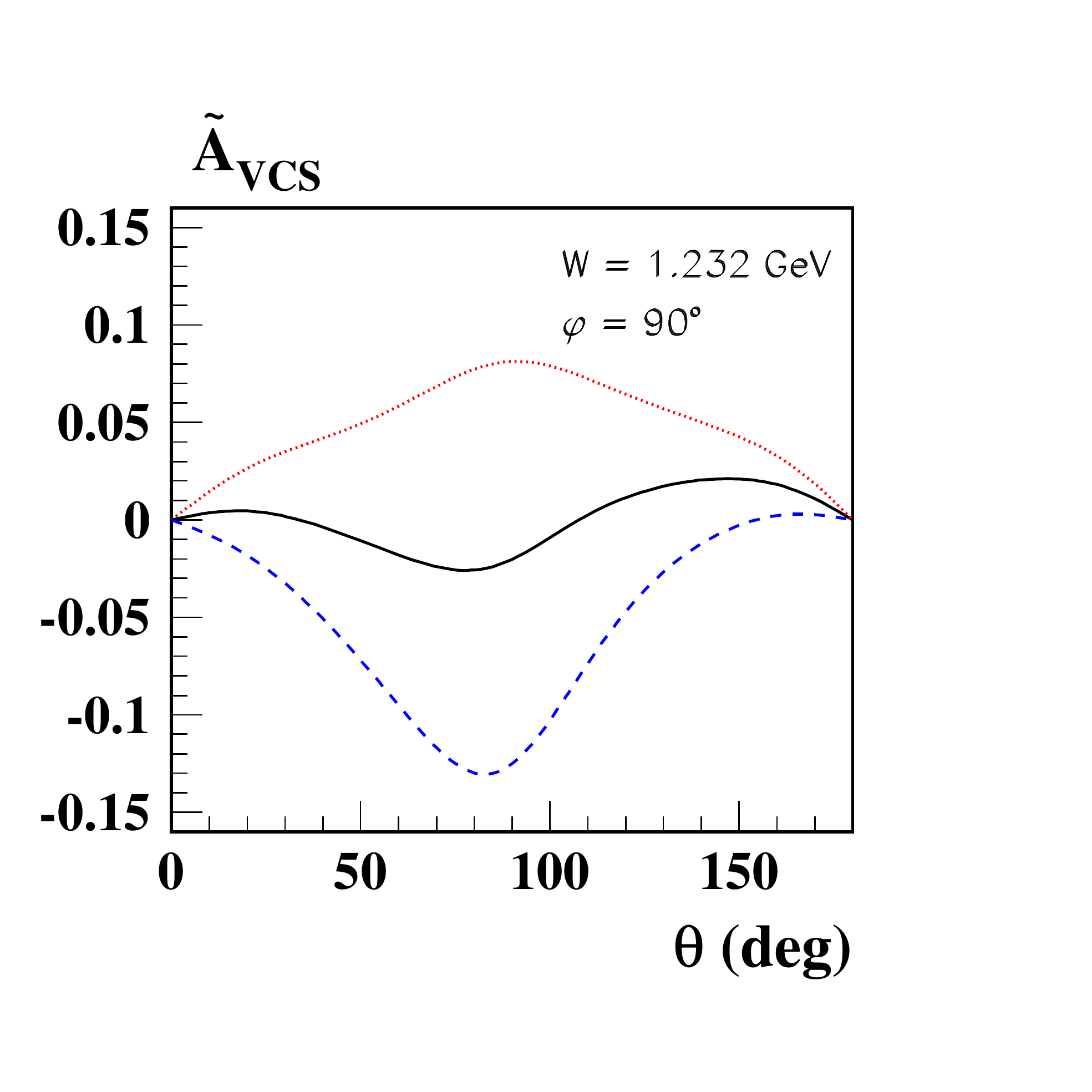}
 \includegraphics[width=0.42\textwidth]{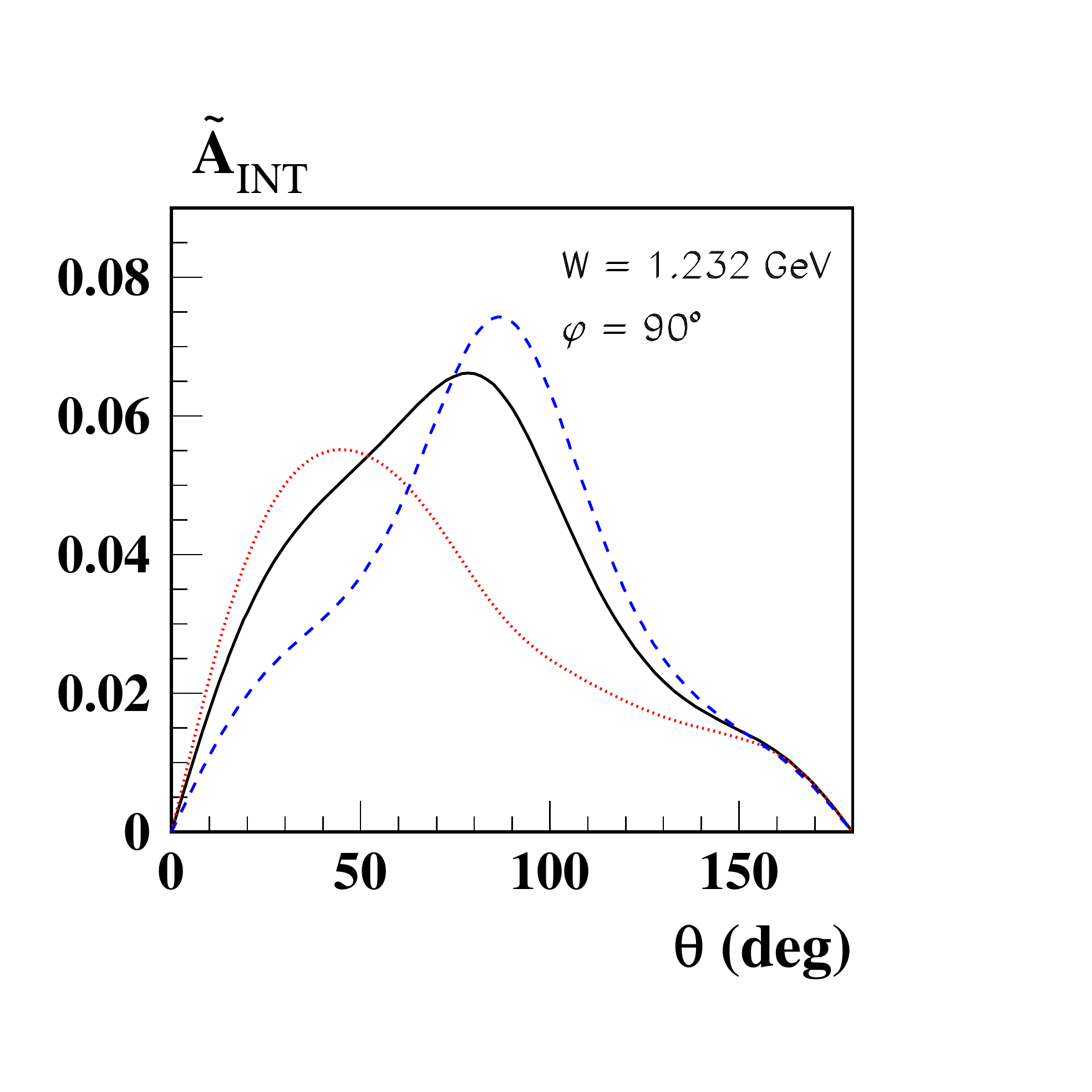}
 \end{center}
    \caption{The asymmetry combinations 
    $\tilde A_{\mathrm{VCS}}$ of Eq.~(\ref{eq:combplus}) and $\tilde A_{\mathrm{INT}}$ of Eq.~(\ref{eq:combminus}),  as function of the photon scattering angle for out-of-plane kinematics, for $Q^2=0.65$ GeV$^2$ and the c.m. energy $W=1.17$ GeV (upper panels) and $W=1.232$ GeV (lower panels). The results are for $\Lambda_\alpha=\Lambda_\beta=0.5$ GeV (red dotted curves), $\Lambda_\alpha=\Lambda_\beta=0.7$ (black solid curves) and $\Lambda_\alpha=\Lambda_\beta=0.9$ GeV (blue dashed curves). }
\label{fig:7}       
\end{figure*}

We study the beam-spin asymmetries for electron and positron beams in the same out-of-plane kinematics as in 
Figs.~\ref{fig:3} and \ref{fig:4}, 
for $\phi = 45^{\rm o}$ and $\phi = 90^{\rm o}$, both 
for $Q^2=0.43$ GeV$^2$ in Fig.~\ref{fig:5}, 
and for $Q^2=0.65$ GeV$^2$ in Fig.~\ref{fig:6}. 
One notices that these asymmetries have a large 
sensitivity to the GP, which is maximal 
for $\phi = 90^{\rm o}$ and for kinematics corresponding with the excitation of the $\Delta(1232)$ resonance.
To better understand this sensitivities, we consider the 
observables $ \tilde A_{\mathrm{VCS}}$ and $\tilde A_{\mathrm{INT}}$ defined in Eqs.~(\ref{eq:combplus}) and (\ref{eq:combminus}) respectively, which involve the 
beam-charge asymmetry and the beam-spin asymmetries for both electron and positron beams. They allow to separate the contributions from the $d\tilde{\sigma}_{\mathrm{VCS}}$ and $d\tilde{\sigma}_{\mathrm{INT}}$ cross section contributions. 
We show these observables in Fig.~\ref{fig:7} for $\phi = 90^{\rm o}$ and $Q^2 = 0.65$~GeV$^2$, for both $W = 1.17$~GeV and $W = 1.232$~GeV. 

We notice from Fig.~\ref{fig:7} that the combination 
$ \tilde A_{\mathrm{VCS}}$, which mainly stems from the 
interference product 
between the imaginary part of the transverse VCS amplitude due to $\Delta(1232)$ excitation and the real part of the longitudinal VCS amplitude, shows a large sensitivity to the GPs. This sensitivity mainly comes from the GP $\alpha_{\mathrm{E1}}(Q^2)$. The asymmetry combination $ \tilde A_{\mathrm{VCS}}$ thus provides an independent way to experimentally access the electric dipole GP. Comparing the $ \tilde A_{\mathrm{VCS}}$ results for 
$W = 1.17$~GeV and $W = 1.232$~GeV, we see that the observable reaches the largest values at $W = 1.232$~GeV as the imaginary part of the VCS amplitude is maximized on the $\Delta(1232)$ resonance position. 

On the other hand, we also notice from Fig.~\ref{fig:7} that 
the combination $ \tilde A_{\mathrm{INT}}$ is much less sensitive to the GPs, especially on the $\Delta(1232)$ resonance position. 
This can be understood as $d\tilde{\sigma}_{\mathrm{INT}}$ measures the interference product between the BH amplitude and the imaginary part of the VCS amplitude, which does not depend on the GPs at all. The remaining GP sensitivity which one sees on Fig.~\ref{fig:7} for $ \tilde A_{\mathrm{INT}}$ results from the denominator contribution $d\sigma_{\mathrm{VCS}}$ in Eq.~(\ref{eq:combminus}). The observable $ \tilde A_{\mathrm{INT}}$ provides however a test for the unitary input in the dispersive formalism.

\section{Conclusions}

In this work we have presented an impact study for an experimental program to measure the VCS process $l p \rightarrow l p \gamma$ with positron beams at Jefferson Lab. Using a dispersive formalism to extract the generalized polarizabilities (GPs) from the VCS observables, we have shown that the beam-charge asymmetry has a sizeable sensitivity to the GPs in the $\Delta(1232)$ resonance region, when going to out-of-plane kinematics. Furthermore, by measuring the beam-spin asymmetry for a positron beam and combining it with the corresponding measurement with an electron beam allows to define observables which separate the cross section terms resulting from the real part of the longitudinal VCS amplitude, which has a strong sensitivity to the electric dipole GP $\alpha_{\mathrm{E1}}(Q^2)$, and the term proportional to the imaginary part of the VCS amplitude, which provides a cross-check on the unitary input in the dispersive formalism.

\begin{acknowledgements}

The work of M.V. is supported by the Deutsche Forschungsgemeinschaft (DFG, German Research Foundation), in part through the Collaborative Research Center [The Low-Energy Frontier of the Standard Model, Projektnummer 204404729 - SFB 1044], and in part through the Cluster of Excellence [Precision Physics, Fundamental Interactions, and Structure of Matter] (PRISMA$^+$ EXC 2118/1) within the German Excellence Strategy (Project ID 39083149).

\end{acknowledgements}

\bibliographystyle{spphys}  
\bibliography{vcsbiblio.bib}   %

\end{document}